\documentclass[prx,aps,twocolumn,nofootinbib]{revtex4}

\usepackage{bm}
\usepackage{subfigure}
\usepackage{youngtab}
\usepackage{amsmath,amssymb}
\usepackage{color,multirow}
\usepackage{graphicx,nicefrac}
\usepackage{epstopdf}
\usepackage[samesize]{cancel}

\usepackage{bm}
\usepackage{hyperref}
\usepackage{xcolor}

\begin{document}
\title{Chiral topologically ordered insulating phases in arrays of interacting integer quantum Hall islands}

\author{Hiromi Ebisu$^1$, Rohit R. Kalloor$^1$, Alexei M. Tsvelik$^2$, Yuval Oreg$^1$
}
\affiliation{
$^1$~Weizmann Institute of Science, Rehovot, Israel 76100\\
$^2$~Condensed Matter Physics and Materials Science Division,
Brookhaven National Laboratory, Upton, NY 11973-5000, USA
}
\date{\today}
\begin{abstract}
We study networks of Coulomb-blockaded integer quantum Hall islands with even fillings $\nu=2k$ ($k$ being an integer), including cases with $2k$ layers each of $\nu=1$ fillings. Allowing only spin-current interactions between the islands (i.e., without any charge transfer), we obtain solvable models leading to a rich set of insulating $SU(2)_k$ topologically ordered phases. The case with $k=1$  is dual to the Kalmeyer-Laughlin phase, $k=2$ to Kitaev's chiral spin liquid and the Moore-Read state, and  $k=3$  contains a Fibonacci anyon that may be utilized for universal topological quantum computation. Additionally, we show how the $SU(2)_k$ topological phases may be obtained also in an array of islands with $\nu=2k$ integer quantum Hall states and critical spin chains in a checkerboard pattern. 
The array and checkerboard constructions gap out the charge modes and additional ``flavor" modes by virtue of their geometry. Furthermore, we find that a fine tuning of the system parameter is not needed in the checkerboard configuration and the $\nu=2$ case. 
We also discuss
their bulk excitations, and show that their thermal Hall conductance is universal, reflecting the central charge $c=3k/(k+2)$ of the chiral edge modes. 
\end{abstract}
\pacs{pacs}
\maketitle

\thispagestyle{empty}
\section{Introduction}
Topologically ordered phases of matter~\cite{Tsui,Wen1990,Kitaev2003} have made a tremendous impact on condensed matter communities. One of the intriguing properties of these phases is that they have exotic fractionalized excitations~\cite{Laughlin,Wilczek1982} - anyons. Generally, anyons fall into two categories - Abelian  and non-Abelian~\cite{Moore1991,Wen1991}. Exchanging two Abelian anyons gives a $U(1)$ phase factor, $e^{i\theta}$ with $\theta\neq 0,\pi$. They exhibit fractional statistics neither bosonic or fermionic. In contrast, the statistics of non-Abelian anyons is described by a square matrix acting on degenerate eigenstates. These matrices in general do not commute, forming a non-Abelian group - hence the term non-Abelian anyons. Non-Abelian anyons may find  applications in quantum computation~\cite{Freedman2003,Nayak2008}.\par

As far as realizations are concerned, it is known theoretically that both Abelian and non-Abelian topological phases exist in fractional quantum Hall (FQH) fluids with various filing fractions. 
One example of an Abelian topological phase is the FQH state at filling fraction $\nu=1/3$ which has anyonic quasi-hole excitations with fractional charge $e/3$~\cite{De1998,Saminadayar1997} and fractional statistics $\theta=\pi/3$.  
The Moore-Read state~\cite{Moore1991} at filling fraction $\nu=5/2$, is a paramount example of the non-Abelian topological phase, and is known to have the so-called Ising anyon. Experimental signatures of the Moore-Read state have been observed recently by investigating thermal current~\cite{banerjee2018observation}. However, from the quantum information perspective,  the Ising anyon has been shown to be non-universal and hence insufficient for the realization of universal topological quantum computers~\cite{Freedman2006,Nayak2008}. Therefore, it is desirable to realize non-Abelian topological phases beyond the Moore-Read state.

In this paper, we explore the emergence of rich topological phases in arrays of interacting integer quantum Hall (IQH) islands with an even integer filing fraction $\nu=2k$ ($k$ being an integer). The topologically ordered phases we obtain are the so-called $SU(2)_k$ topological phases, where the bulk is described by $SU(2)$ Chern-Simons topological gauge theory~\cite{CS} with level $k$. Accordingly, due to the bulk-edge correspondence, the edge theory is given by $SU(2)_k$ Wess-Zumino-Novikov-Witten (WZNW) conformal field theory (CFT)~\cite{Wess1971,Novikov1981,Witten1983,Witten1984} with central charge $c=3k/(k+2)$. Our proposal is based on the network construction, originally suggested by Chalker and Coddington~\cite{Chalker1988} and recently updated by Hu and Kane~\cite{Hu2018} in the context of interacting $p$-wave superconductors. 

The $SU(2)_1$ topological phase is intriguing in its own right as it is identified with the Kalmeyer-Laughlin (KL) state~\cite{Kalmeyer1987}, which is one of the spin liquid phases known to have deconfined fractional spin excitations. Essentially, this phase is topologically equivalent to the bosonic FQH state with filling fraction $\nu=1/2$, with only one type of Abelian anyon - namely the semion with fractional statistics $\theta=\pi/2$ - as an excitation. Notice that the semion is neutral in the spin liquid version of the KL state while in the bosonic $\nu=1/2$ FQH state, it carries a fractional charge $e/2$.  
The $SU(2)_2$ phase is dual to the Moore-Read state and the Kitaev chiral spin liquid phase~\cite{Kitaev2006} containing the Ising anyon. 
The importance of the $SU(2)_3$ topological phase is derived from the fact that  it has a special kind of non-Abelian anyon - the Fibonacci anyon~\cite{Freedman2002,Freedman2003,Feiguin2007}. Its statistics is defined by the fusion rule  $\tau\times \tau=I+\tau$, where $\tau$ is the anyon and $I$ is the trivial particle. \par
In view of the network construction, the basic principle to realize our phases is to gap out the $SU(2)_k$ sector. To this end, we will demonstrate two configurations that stabilize our phases. The first one is composed of networks of Coulomb-blockaded $\nu=2k$ IQH islands where adjacent islands are interacted by the $SU(2)$ current consisting of the chiral edge modes of the $\nu=2k$ IQH state. In the second geometry, IQH islands and insulating islands hosting a critical spin chain on the perimeter are placed in a checker board pattern. The interaction between adjacent IQH-spin islands is comprised of the $SU(2)$ current of the IQH island and the spin current of the critical spin chain. 
\par
We identify several ways to excite anyonic quasi-particles in the bulk. In the case of IQH islands, an anyon arises as a soliton in the gapped area between adjacent islands, or by introducing magnetic fluxes within the islands. Especially,  the soliton is identified as the semion for $k=1$, the Ising anyon for $k=2$, and  the Fibonacci anyon for $k=3$. In the case with spin chains, local spin excitations are associated with the anyons.

The outline of this paper is as follows. 
In Sec.~\ref{model}, we introduce a model consisting of an array of Coulomb-blockaded islands at filling $\nu=2k$ to construct the $SU(2)_k$ topological phase and its conjugate $U(1) \times SU(k)_2$ phase (when the array is embedded in an $\nu=2k$ IQH state).
We assume that the islands have spin degenerate states and due to the Coulomb blockade, charge cannot be transferred between the islands so that the whole system is insulating.  In addition, we introduce spin current interactions between the islands.
The introduction of  the spin current interactions only demands fine tuning that may be challenging  in real systems, but makes the model solvable. The phases we obtain are gapped in the bulk and contain chiral edge modes; we expect therefore that an introduction of perturbations smaller than the bulk gap will not lead to a phase transition and will not modify the universal properties of the topologically ordered phase. 

Sec.~\ref{nu2kl} describes the KL state, which has in our construction a neutral mode that propagates along the edge, and accordingly, thermal Hall conductance $\kappa = 1$ (in units of $\kappa_0 = \frac{\pi^2 k_B^2}{3 h} T$; with $T$ being the temperature, $k_B$ the Boltzmann constant, and $h$  Planck's constant). In the conjugate phase, a sole $U(1)$ charge mode (with Hall conductance $2 e^2/h$ and $\kappa=1$) propagates along the edge.
 \,\,\,
In Sec.~\ref{Maj}, we briefly discuss the $SU(2)_2$ phase which is dual to both the Moore-Read and the Kitaev spin liquid states, and contains the Ising anyons. The insulating phase has $\kappa=3/2$; its conjugate phase has a charge mode with Hall conductance equal to $4 e^2/h$, and together with the conjugate neutral modes, yields $\kappa=5/2$.  \,\,\,
Sec.~\ref{Fib} describes the $SU(2)_3$ phase which contains four types of excitations, with one of them being the Fibonacci anyon. In this phase, $\kappa = \left. {3 k}/{(k+2)}\right|_{k=3} = 6/5$, while in the conjugate phase the Hall conductance is $6e^2/h$ and $\kappa= 2k-{3 k}/{(k+2)}= \left.k {(2k+1)}/{(k+2)}\right|_{k=3} = 
 21/5 $.\,\,\, 
In Sec~\ref{fibph}, we will demonstrate that a combination of the $SU(2)_1$ and $SU(2)_3$ phases may lead to the Fibonacci topological phase~\cite{Mong2014,Hu2018}, a topological phase having only the trivial and the Fibonacci anyon as excitations.
 We show that the  Fibonacci topological phase is stabilized if, in addition to the pure spin current interactions, anyons in the $SU(2)_1 \times SU(2)_3$ phase condense.
 

In Sec.~\ref{alt}, we present an alternative construction of the $SU(2)_k$ topological phase by introducing a checker board  alternating between IQH islands and $SU(2)_k$ critical spin chains on the perimeter of a vacuum. In this construction, the pure spin current interaction is more natural.  
This IQH-spin chain model allows us to discuss the renormalization group flow of the system to its strong coupling limit, and show, without referring to topological arguments, why it is stable for odd values of $k$.

In Sec.~\ref{sec:experiment}, we comment on experimental consequences and realizations of the topological phases, and finally, Sec.~\ref{conclusion} is devoted to conclusions. Technical details are relegated to appendices. 

\begin{figure*}[t]
\subfigure[]{%
		\includegraphics[clip, width=0.6\columnwidth]{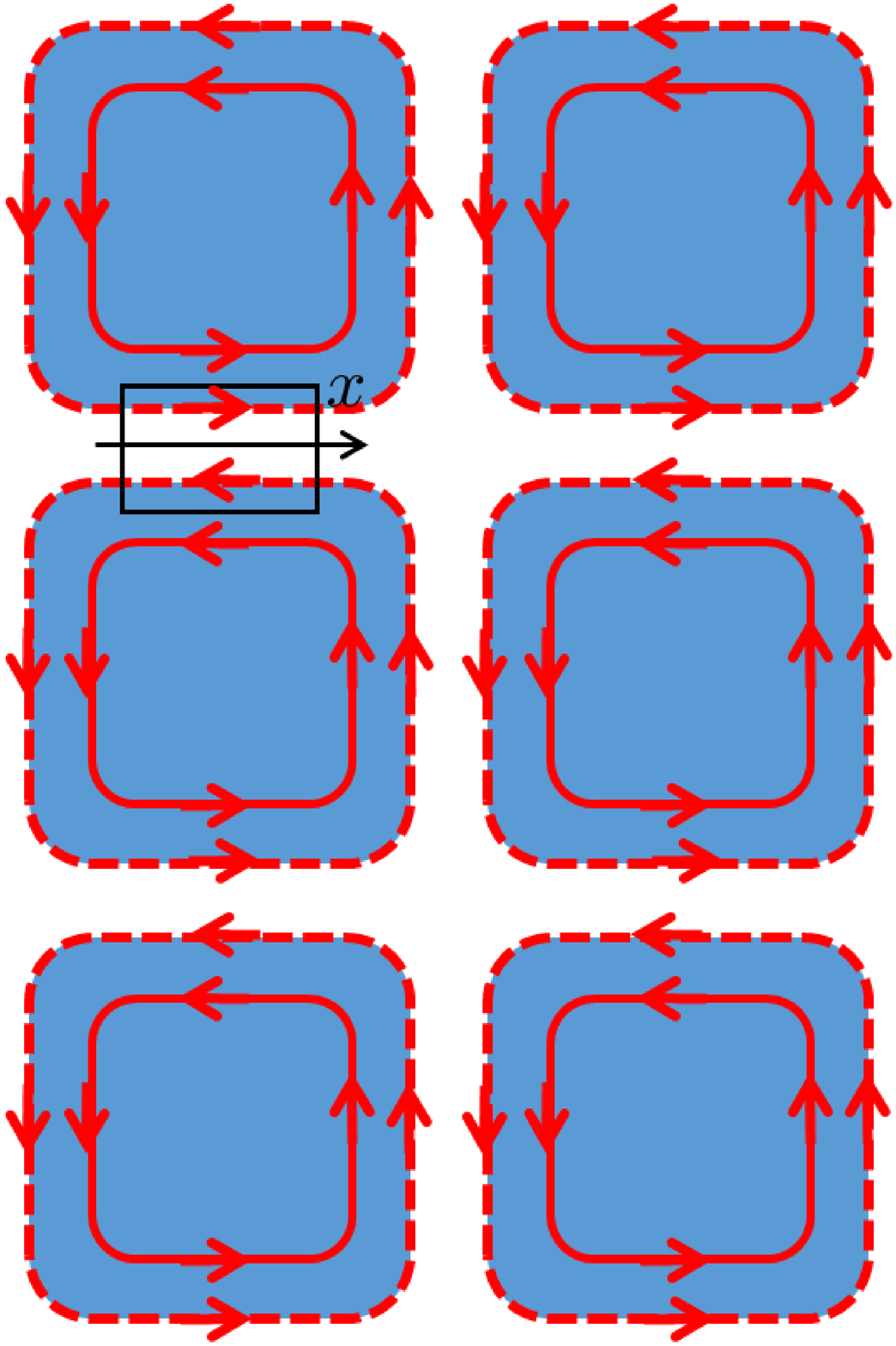}}
\hspace{1cm}
\subfigure[]{%
		\includegraphics[clip, width=0.6\columnwidth]{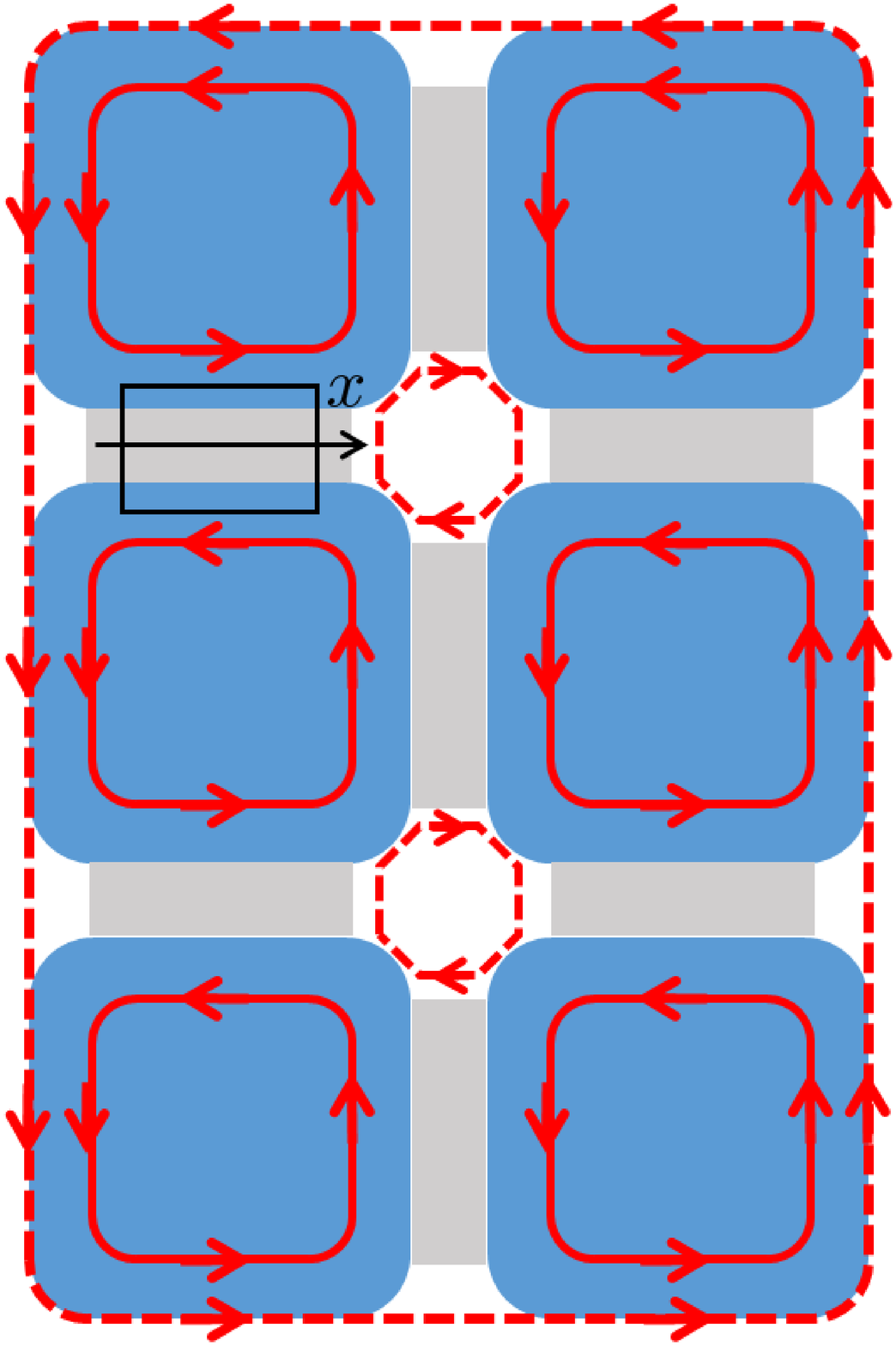}}%
\hspace{0cm}
\subfigure[]{%
		\includegraphics[clip, width=0.71\columnwidth]{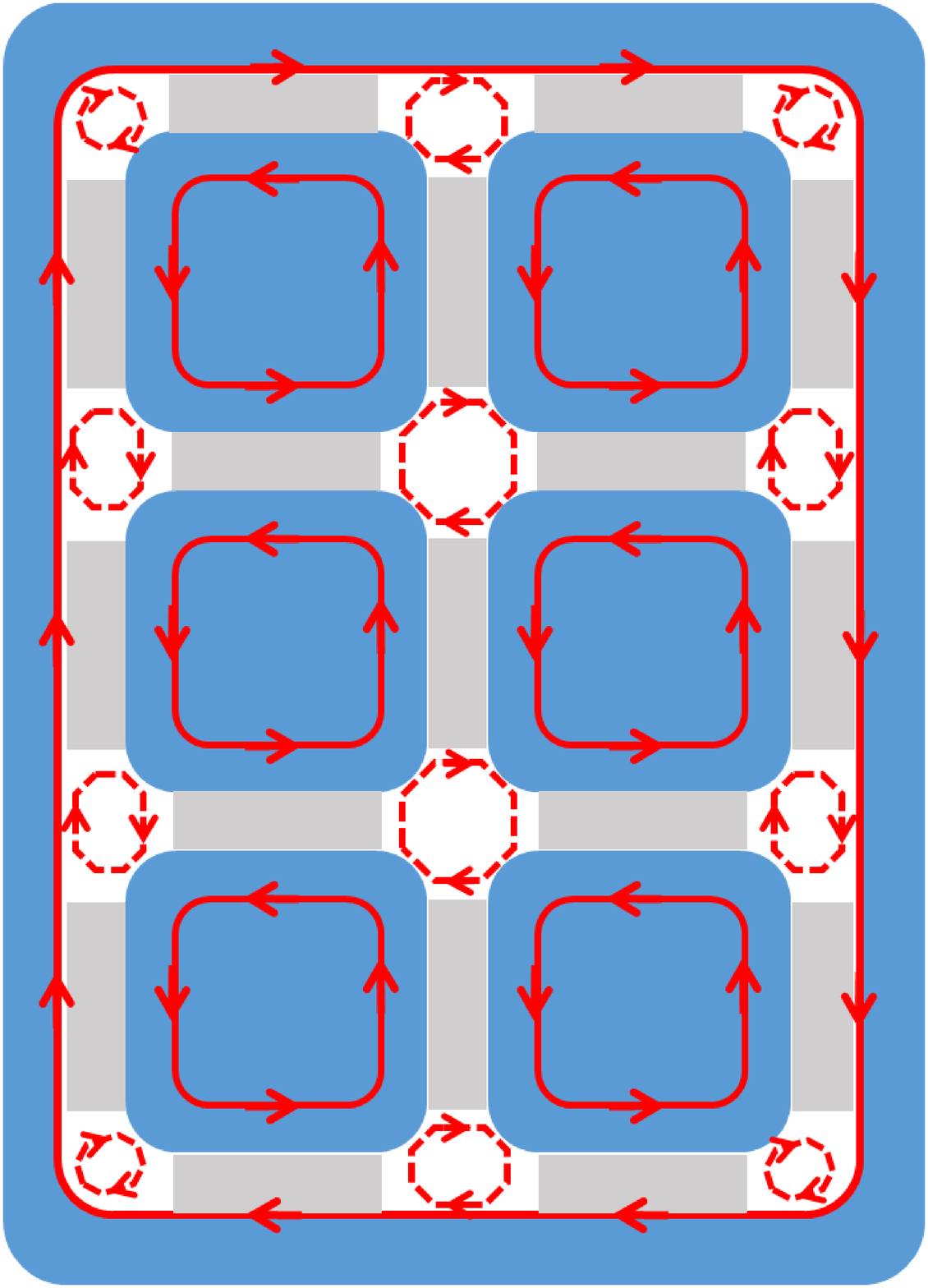}}%
	\caption{(a) Integer quantum Hall (IQH) islands with filling fraction $\nu=2k$ which has $2k$ chiral edge modes moving in the counterclockwise direction. These modes are decomposed into the
$U(1)\oplus SU(k)_2$ sectors marked by the bold red line, and the $SU(2)_k$ sector depicted by the dashed red line.	
(b) Networks of the \textit{interacting} IQH islands -- spin current interactions are turned on (only) inside the grey areas. As a result, the charge and flavor characterized by the $U(1)\oplus SU(k)_2$ sectors (bold red line) are confined to each IQH island, whereas the $SU(2)_k$ sector (dashed red line)  propagates inside the vacuum (white) area or along the entire edge of the system, yielding the $SU(2)_k$ topological phase. (c) A configuration to realize a conjugate phase of the $SU(2)_k$ topological phase, consisting of the network of IQH islands that interfaces with a large IQH system with filling fraction $\nu=2k$.  The red dashed and bold lines represent the $SU(2)_k$ and $U(1)\oplus SU(k)_2$ sectors respectively. Notice that when $k=1$, the flavor degree of freedom is absent, hence the bold red line represents the $U(1)$ sector. }\label{check}
	
\end{figure*}
\section{model}\label{model}
In this section, we describe  a network model to construct  different states that are dual to the KL state, the Kitaev honeycomb spin liquid (which is dual to the Moore-Read state), 
and the $SU(2)_3$ topological phase hosting the Fibonacci anyons. Furthermore, we discuss a way to obtain the Fibonacci phase. 

\subsection{ \texorpdfstring{$\nu=2$}{Lg}: The Kalmeyer-Laughlin state}\label{nu2kl}
Let us start with the simplest phase generated by our construction, that is, the KL state. This phase has a chiral edge mode characterized by the $SU(2)_1$ WZNW CFT with central charge $c=1$.
We prepare IQH islands in a square shape at filling fraction $\nu=2$, with two chiral edge modes propagating in the counter-clockwise direction and consider networks of these IQH islands as portrayed in Fig. \ref{check}(a). \par

 Each IQH island has two chiral edge modes which respect $U(2)$ symmetry. The corresponding chiral Hamiltonian  can be decomposed as $U(2)=U(1)+SU(2)$, which is reminiscent of the decomposition into charge and neutral modes in a FQH system~\cite{KaneFisher}. In the context of the Tomonaga-Luttinger liquid,  this decomposition can be interpreted as spin-charge separation~\cite{Tomonaga,Luttinger}. The precise form of the decomposition is given by the  conformal embedding, which reads as
\begin{equation}
U(2)_1=SU(2)_1\oplus U(1),\label{emd}
\end{equation}
where the number in the subscript represents the level of the WZNW CFT.  We refer the interested reader unfamiliar to a review article on conformal embedding~\cite{james2018non}. The embedding (\ref{emd}) is suggestive of  the KL state, as on the right hand side (r.h.s) of Eq.~(\ref{emd}), the $SU(2)_1$ sector appears, which is exactly what characterizes the KL state. \par
To proceed further, we introduce an interaction between adjacent IQH islands. In this Section we will not say much about microscopic realization of such interaction. Now we concern ourselves only with the principle.
One possible  microscopic realization is discussed later in Sec.~\ref{alt}. (See also Appendix.~\ref{bosonization} for discussion of a bosonized description of the interaction.)  
 The interaction acts only in areas where the islands are close to each other, as
inside the black frame as shown in Figs. \ref{check}(a) and \ref{check}(b). In this area, there are two pairs of counter-propagating modes. Denoting $\psi_{R,\alpha}$ ($\psi_{L,\alpha}$) as the Dirac field corresponding to edge modes of the top (bottom) IQH island inside the black frame, with $\alpha=1,2$ or interchangeably $\alpha=\uparrow,\downarrow$ ,
we define the following currents
\begin{eqnarray}
J^x_{R/L}&=&\frac{1}{2}(\psi^{\dagger}_{R/L,1}\psi_{R/L,2}+\psi^{\dagger}_{R/L,2}\psi_{R/L,1}),\nonumber\\
J^y_{R/L}&=&\frac{1}{2}(-i\psi^{\dagger}_{R/L,1}\psi_{R/L,2}+i\psi^{\dagger}_{R/L,2}\psi_{R/L,1}),\nonumber\\
J^z_{R/L}&=&\frac{1}{2}(\psi^{\dagger}_{R/L,1}\psi_{R/L,1}-\psi^{\dagger}_{R/L,2}\psi_{R/L,2}).\nonumber\label{curren0}
\end{eqnarray}
These currents have an $SU(2)$ symmetry, allowing us to write them in a more compact form as 
\begin{equation}
J^a_{R/L}=\sum_{\alpha,\beta=1,2}\psi_{R/L,\alpha}^{\dagger}\frac{\sigma^a_{\alpha\beta}}{2}\psi_{R/L,\beta}\;(a=x,y,z)\label{current01}
\end{equation} 
with $\sigma^a_{\alpha,\beta}$ being the $SU(2)$ generators.
The Hamiltonian describing the modes inside the black frame is therefore given by
\begin{eqnarray}
H_{\text{2}}=&&\int dx\sum_{\alpha}v(i\psi^{\dagger}_{R,\alpha}\partial_x\psi_{R,\alpha}-i\psi^{\dagger}_{L,\alpha}\partial_x\psi_{L,\alpha})\nonumber\\
&+&\sum_{a=x,y,z}\lambda_2J_{R}^aJ_{L}^a,\label{interaction}
\end{eqnarray}
where $v$ is the velocity of the Dirac fields,  $x$ is the one-dimensional coordinate in the frame, and $\lambda_2$ is the coupling constant.
 As we discuss in Sec.~\ref{sec:experiment} (and Fig.~\ref{dl}), such an interaction naturally emerges in  networks of double layer of $\nu=1$ states.
We assume that our network has interaction ~(\ref{interaction}) taking part between all adjacent islands, that is in the areas with grey shading in Fig.~\ref{check}(b) (and only in these areas).

At $\lambda_2 >0$ the current-current interaction given in Eq.~(\ref{interaction}) is marginally relevant and gaps out the $SU(2)_1$ sector in Eq.~(\ref{emd}), yielding the desired KL state, $i.e.,$ the $SU(2)_1$ topological phase. Indeed, the edge mode of the ungapped sector on each island [the $U(1)$ sector in Eq.~\eqref{emd}]  passes through the interacting area, but the $SU(2)_1$ mode bounces off. As a consequence, the $U(1)$ sectors remain confined to each IQH island [see red bold lines in Fig.~\ref{check}(b)]. This feature constitutes a great advantage of the network array construction over the wire construction one: one does not need to introduce additional interactions to gap out the charge mode. On the other hand, the edge modes  of the gapped $SU(2)_1$ sector are not transmitted through the interaction areas and hence become confined to the vacuum regions [white regions in Fig.~\ref{check}(b)]. However, as is clear from Fig.~\ref{check}(b), one chiral mode is free to  propagate  along the entire edge of the system [red dashed line in Fig.~\ref{check}(b)], which results in the $SU(2)_1$ topological phase -- the topological phase with a neutral chiral edge mode described by the $SU(2)_1$ WZNW CFT with central charge $c=1$. The suggested mechanism imposes restrictions on the size of the islands. Namely, the size of the interacting area must be larger than the correlation length $\xi$ of the spin sector. 
When the current-current interaction is isotropic, which is the case for Hamiltonian ~(\ref{interaction}), we have  $\xi \sim l_B\exp(\pi v/\lambda_2), ~ (l_B $ is the magnetic length which plays the role of the ultraviolet cut-off).  
\par

At $k =1$ there is only one nontrivial anyonic excitation $s$ with conformal weight $1/4$, corresponding to a primary field of the $SU(2)_1$ WZNW CFT, and fusion rule $s\times s=I$. The $s$ anyon is nothing but the semion described above. Its  fractional statistics can be obtained from the conformal weight and the fusion rule, giving $\theta=\pi/2$. \par
The semion excitation can be seen in the regions in which the spin sector is gapped by the interaction in Eq.~(\ref{interaction}) [the areas with grey color in Fig.~\ref{check}(b)]. Deferring the details to the Appendix.~\ref{bosonization}, it turns out that the gapped theory between adjacent IQH islands has two-fold ground state degeneracy, which is intuitively understood as two ferromagnetic ground state configurations of spins where all  spins point up or down. A kink trapped at the interface between domains of such configurations, is characterized by conformal weight $1/4$. This implies that the kink bound at the interface can be regarded as the semion.  \par

The excitation of the semion also arises by inducing a magnetic flux in an IQH island. 
In order to discuss a physically legitimate implementation to have such a flux, we let the $\nu=2$ IQH island be comprised of two layers of quantum Hall states each of which has filling $\nu=1$. In addition, we assume that a gate can alter the number of electrons in one of the layers and deplete the charge. Since the filling fraction of the layer we are tuning is $\nu=1$, the ratio between the number of magnetic flux and the electrons is unity, implying a charge depletion is associated with introduction of  a unit of magnetic flux $h/e$ . In what follows, we dub such a magnetic flux as a ``$h/e$ vortex" characterized by a conformal weight of an electron excitation $1/2$.
From the conformal embedding in Eq.~(\ref{emd}), this excitation consists of spin [$SU(2)_1$] and charge [$U(1)$] degrees of freedom, which is described by~\cite{NACULICH1990}
\begin{equation}
1/2=1/4+1/4.\label{semi}
\end{equation}
The first term corresponds to the conformal weight of the semion in the $SU(2)_1$ sector, and the second term comes from the conformal weight of the charged excitation of the $U(1)$ sector.
Since the resulting phase of our model is given by $SU(2)_1=\frac{U(2)_1}{U(1)}$, the $U(1)$ sector is suppressed, (physically, corresponding to the fact that we don't admit any charge transfer between the islands) allowing us to omit the second term in Eq.~(\ref{semi}). Therefore,
the vortex excitation behaves as the semion. In the analogy with the fact that a $h/(2e)$ vortex trapping the Ising anyon, a.k.a. the Majorana zero mode in the bulk of a $p$-wave superconductor forces the edge mode to carry the Majorana zero mode~\cite{corbino}, we expect that in the KL state, the edge mode has the semion in accordance with the semion excitation in the bulk induced by a vortex.

\par
In the same spirit of the work by Hu and Kane~\cite{Hu2018}, we can construct a ``conjugate phase" of the KL state - in our case, the $U(1)$ topological phase. The $U(1)$ topological phase and the KL state are conjugate to each other in the sense of the conformal embedding. 
To generate the conjugate phase, we surround our network of IQH islands by a large $\nu=2$ IQH system which has two chiral edge modes propagating in clockwise direction, see Fig.~\ref{check}(c). We further introduce interactions between the large IQH system and the networks of the IQH islands in the same form as the second line of Eq.~(\ref{interaction}). From the arguments similar to the ones  given above, we obtain a topological phase which has a $U(1)$ chiral edge mode with central charge $c=1$ propagating in the clockwise direction along the interface between the large IQH system and the networks of IQH islands~\cite{conjugatephase}.
Notice that 
the bulk excitation in the conjugate phase is described by the same neutral semion as the KL state, as configurations of the bulk are the same for both phases - the networks of $\nu=2$ IQH islands shown in Fig.~\ref{check}(b). Hence, a neutral excitation in the bulk of the conjugate phase shows a discrepancy with the edge excitation which carries a charge. Such a discrepancy is resolved by noting that the multiplication of a semion in the bulk and a $h/e$ vortex in the large IQH that surrounds the networks corresponds to the edge excitation described by the second term of the r.h.s of Eq.~(\ref{semi}) which represents a charged excitation in the $U(1)$ sector with conformal weight $1/4$ carrying a charge $e$. We interpret this multiplication of the excitations from the bulk and outside the bulk as the excitation of the conjugate phase. \par

The physical difference between these two phases can be seen by measuring charge conductance. For the KL state, the edge mode is neutral and thus, the charge conductance is zero. On the other hand, for the conjugate phase of the KL state, charge conductance is two (in units of $e^2/h$). The neutral mode of the KL state can be probed by measurement of thermal conductance.

\subsection{\texorpdfstring{$\nu =4$: $SU(2)_2$}{Lg} topological phase}\label{Maj}
Let us now move on to the construction of the $SU(2)_2$ topological phase. In doing this we will closely follow the method of the previous subsection. We consider an IQH island with filling fraction $\nu=4$ in a square shape. We introduce a geometry of networks of the IQH island as depicted in Fig.~\ref{check}(a). Similar to the previous subsection, introducing the Dirac fields by $\psi_{R/L,\alpha,i}$ ($\alpha=1,2$, $i=1,2$), we write  the Hamiltonian between adjacent islands as
\begin{eqnarray}
H_{\text{4}}=&&\int dx\sum_{\alpha,i}v(i\psi^{\dagger}_{R,\alpha,i}\partial_x\psi_{R,\alpha,i}-i\psi^{\dagger}_{L,\alpha,i}\partial_x\psi_{L,\alpha,i})\nonumber\\
&+&\sum_{a=x,y,z}\lambda_4J_{R}^aJ_{L}^a\label{interaction6}
\end{eqnarray}
with $SU(2)$ current 
\begin{equation}
J^a_{R/L}=\sum_{\substack{\alpha,\beta=1,2\\i=1,2}}\psi_{R/L,\alpha,i}^{\dagger}\frac{\sigma^a_{\alpha\beta}}{2}\psi_{R/L,\beta,i}\;(a=x,y,z).\label{current06}
\end{equation} 
Notice that each IQH island has four chiral edge modes; accordingly, compared with the previous case, we have a new index $i=1,2$ in addition to $\alpha=1,2$ to denote the Dirac field.   \par
As opposed to the previous case, we will exploit a more complicated conformal embedding with regard to the four edge modes of the IQH island: 
\begin{equation}
U(4)_1=SU(2)_2\oplus U(1)\oplus SU(2)_2\label{emb6}.
\end{equation} 
The first, second and third sectors correspond to the spin, charge, and flavor degrees of freedom, respectively. Notice that the spin and flavor sectors are characterized by the same symmetry, namely $SU(2)_2$, which is special for the case  $k=2$.
Interaction in Eq.~(\ref{interaction6}) is not the most general current-current interaction, but is tailored to exploit this embedding, which constitutes a potential problem for practical realizations. A way  to resolve this difficulty is discussed in the next section. Since the spin $SU(2)_2$ currents commute with the part of the Hamiltonian describing the other sectors, the charge and flavor sectors corresponding to the second and third terms in Eq.~(\ref{emb6}) remain unaffected. The current-current interaction in Eq.~(\ref{interaction6}) gaps out only the spin $SU(2)_2$ sector, forcing the edge modes of the $U(1)\times SU(2)_2$ sector to be confined within each IQH island; the $SU(2)_2$ edge modes propagate inside the vacuum areas or along the entire edge of the sample. This
results in the $SU(2)_2$ topological phase with central charge $c=3/2$,
corresponding to three chiral Majorana fermions. \par
In this phase, there are
three types of excitations, $I$, $\psi$, $\sigma$ with fusion rules $\psi\times\psi=I$, $\psi\times\sigma=\sigma$, $\sigma\times\sigma=I+\psi$. This phase behaves as the anti-Pfaffian state~\cite{levin2007particle,lee2007particle}, one of a candidate state of a FQH state  at $\nu=5/2$. Even though the fusion rules of the anyonic excitations and the central charge are identical in these two phases, there are several differences; the filling fraction of the anti-Pfaffian is $\nu=5/2$, on the other hand, the $SU(2)_2$ topological phase is constructed by $\nu=4$ IQH islands. 
Also, in the anti-Pfaffian state, there are charge modes which propagate along the edge whereas in the $SU(2)_2$ phase, there are only  neutral modes.  \par
We can also construct a conjugate phase similarly to the previous case. In view of Eq.~(\ref{emb6}), the conjugate phase is a non-Abelian topological phase characterized by $U(1)\times SU(2)_2$ WZNW CFT. To see this, consider a geometry of interacting $\nu=4$ IQH islands interacted with a large $\nu=4$ IQH system that surrounds the network [see Fig.~\ref{check}(c)]. 
The interaction has the same form as the terms in the second line in the r.h.s of Eq.~(\ref{interaction6}).
The interaction yields a conjugate phase which has a chiral edge mode described by $U(1)\times SU(2)_2$ WZNW CFT which carries central charge $c=5/2$.\par
Similarly to the previous subsection, the interaction area corresponding to the rectangle marked by grey color in Fig.~\ref{check}(b), may bind the Ising anyon. Defining a bosonic and Majorana field by $\Phi$ and $\chi$, the Lagrangian density corresponding to Hamiltonian in Eq.~(\ref{interaction6}) is written as \cite{tsvelik2014integrable}
\begin{equation}
    \mathcal{L}=\frac{1}{2}(\partial_{\mu}\Phi)^2+Z_2[\chi,\bar{\chi}]-\lambda(e^{i\beta\Phi}\chi\bar{\chi}+\text{H.c.})\label{majorana}
\end{equation}
with $\lambda\sim4\lambda_4$ and $\beta^2=(1+2\lambda_4/4\pi)^{-1}$. 
The first term is a kinetic term of a bosonic field $\Phi$.
The second term describes the critical Majorana theory and the third term represents a mass term. Eq.~(\ref{majorana}) is a field theory description of a generalization of the Majorana chain as the Majorana mass term is dynamical. Spacial modulation of the mass traps a Majorana zero mode, which is identified as the Ising anyon. \par
We can also discuss how vortices in a $\nu=4$ IQH island give rises to the anyon in the $SU(2)_2$ topological phase. In the similar manner as the discussion around Eq.~(\ref{semi}), one can 
envisage that the $\nu=4$ IQH island is consisting of four layers of $\nu=1$ IQH, and show that a $h/e$ vortex may bind the Ising anyon by tuning a gate of one of the layer. 

\subsection{\texorpdfstring{$\nu$=6: $SU(2)_3$}{Lg} topological phase}\label{Fib}
In analogy to the previous subsections \ref{nu2kl} and \ref{Maj}, the construction of the $SU(2)_3$ topological phase is straightforward. Preparing networks of $\nu=6$ IQH islands, and using following conformal embedding
\begin{equation}
    U(6)_1=SU(2)_3\oplus U(1)\oplus SU(3)_2,\label{emb60}
\end{equation}
one obtains the $SU(2)_3$ topological phase by introducing the $SU(2)$ current-current interaction between adjacent islands with Hamiltonian
\begin{eqnarray}
H_{\text{6}}=&&\int dx\sum_{\alpha,i}v(i\psi^{\dagger}_{R,\alpha,i}\partial_x\psi_{R,\alpha,i}-i\psi^{\dagger}_{L,\alpha,i}\partial_x\psi_{L,\alpha,i})\nonumber\\
&+&\sum_{a=x,y,z}\lambda_6J_{R}^aJ_{L}^a\label{interaction6a}
\end{eqnarray}
where the subscript $i$ takes three values, $i.e.,$ $i=1,2,3$. 
\par 

There are four types of anyons in this phase, denoted by $I, X, Y, Z$ corresponding to the primary fields $\phi_i$ ($i=0,1/2,1,3/2$) with conformal weight $h_{\phi_i}=i(i+1)/5$. See also Table.~\ref{labels}.
\begin{table}[htb]
  \begin{tabular}{c|c} 
   $SU(2)_3$ & conformal weight\\ \hline
    $I$ &$0$  \\
    $X$  & $\frac{3}{20}$\\
    $Y$  &$\frac{2}{5}$\\
     $Z$  &$\frac{3}{4}$
  \end{tabular}
  \caption{Labels and conformal weights of four anyons in the $SU(2)_3$ topological phase.}\label{labels}
\end{table}
Recalling the fusion rules of the $SU(2)_3$ CFT, which is provided in the Appendix~\ref{data}, the $Y$ anyon is the Fibonacci anyon with fusion rule $Y\times Y=I+Y$.\par 
From Eq.~(\ref{emb60}), the conjugate phase is the $U(1)\times SU(3)_2$ topological phase.
Up to the trivial $U(1)$ sector, this conjugate phase has six anyonic excitations labeled by $b_0$, $b_3$, $b_{\bar{3}}$, $b_6$, $b_{\bar{6}}$, $b_8$.  
Conformal weights and fusion rules of these excitations are provided in Appendix.~\ref{data}. Based on the data there, $b_6$, $b_{\bar{6}}$ are $Z_3$ parafermions and the $b_8$ anyon is the Fibonacci anyon. \par

\par
We can create fractionalized excitations in the interaction area between adjacent islands depicted  with grey color on Fig.~\ref{check}(c) in full analogy to the way it was done in wire construction \cite{Kane2002}). To see this, we can resort to the semiclassical analysis of our model (\ref{interaction6a}) based on the conformal embedding 
\begin{equation}
    SU(2)_3 = U(1)\oplus Z_3.
\end{equation} 
According to Ref.~[\onlinecite{tsvelik2014integrable}], the Lagrangian density of this model can be recast as  
\begin{eqnarray}
  {\cal L} &=& \frac{1}{2}(1+ \lambda_6/2\pi)(\partial_{\mu}\Phi)^2+Z_3[\Psi,\overline{\Psi}] \nonumber \\ &-&\lambda(e^{i\beta\Phi}\Psi\overline{\Psi}+\text{H.c.}).  \label{parafer}
\end{eqnarray}
where, $\lambda\sim 6\lambda_6$ and $\beta^2=8\pi/3$. The first term is the bosonic kinetic term with $\Phi$ being a bosonic field, the second describes the critical $\mathbb{Z}_3$ parafermion theory, where the $\mathbb{Z}_3$ parafermion is given by $\Psi$, and the third term originates from the interaction of the $x,y$ currents. This term dynamically generates a mass for the parafermions. Moreover, since $\exp(i\beta\Phi)$ field changes sign on the  solitons of $\Phi$-field, the parafermions have zero energy modes bound to the solitons. The solitons with the zero modes attached to them become non-Abelian quasiparticles. This distinguishes them from Laughlin quasiparticles constructed in a somewhat similar way in the wire construction in Ref.~\cite{Kane2002}. 
In Ref.~[\onlinecite{tsvelik2014integrable}], one of the authors has demonstrated that  the zero modes located on the solitons are one dimensional analogues of  the Fibonacchi anyons. It has also been shown that at finite density of solitons the anyons interact so that model~(\ref{parafer}) describes  a generalization of the ``golden chain" \cite{Feiguin2007}. 
Models~(\ref{interaction6a}), and (\ref{parafer}) are closely related to the physics of nucleation of topological liquid ~\cite{gils2009collective,ludwig2011two}. The finite density of solitons can be produced if one  introduces a Zeeman field $g_LH$. Such a field couples to the gradient of the  bosonic field as $g_L H\partial_x\Phi$ which leaves the model integrable. By setting $g_L H$ close to the soliton threshold such  that the density of solitons far exceeds the density of antisolitons, but still remains small  $\exp[ -(M -g_L H)/T]\ll 1$ ($T$ is temperature and $M$ is the soliton mass), one obtains a rarefied gas of Fibonacci anyons \cite{tsvelik2014integrable}. At greater $g_LH$, when the interaction between the anyons becomes important, the system undergoes a crossover to a collective state characterized by the parafermionic Z$_3$ CFT with central charge 4/5.  [Together with $U(1)$ bosonic theory, the total central charge is described by the one of the $SU(2)_3$ WZNW CFT.] The forming of the gapless collective mode closely parallels the nucleation of a topological liquid, where a non-trivial collective gapless mode is realized as the Hilbert space of adjacent non-Abelian anyons is projected to one of the sectors of the fusion channels. This allows us to interpret Eqs.~(\ref{interaction6a}) and (\ref{parafer}) with inclusion of the Zeeman field as concrete models of the nucleation. 

Since model (\ref{interaction6a}) and (\ref{parafer}) is translationally invariant the solitons move. However, they can be trapped by space modulations of their mass coming from modulations of the coupling $\lambda$ and become localized.  
\par
Composite $h/e$ vortices in an IQH island can also give rise to anyons. As mentioned in the previous subsections, such an excitation may occur 
when we assume the IQH island consists of multi-layers of quantum Hall states each of which is $\nu=1$ IQH state. In the present case, suppose the $\nu=6$ IQH island is described by six layers of $\nu=1$ IQH states,
$r h/e$ ($1\leq r\leq 6$) vortex excitation may arise by tuning gates to control the fillings of $r$ layers. 
To clarify the relation between the vortices and anyons in the $SU(2)_3$ topological phase, we need to find how the conformal weight of the vortices is decomposed into the ones in the different commuting sectors, similarly to Eq.~(\ref{semi}). The detailed discussion on this decomposition is provided in the Appendix.~\ref{more}. It turns out that two or four $h/e$ vortices may bind the Fibonacci anyon, $i.e.,$ the $Y$ anyon in the $SU(2)_3$ topological phase.

\subsection{Fibonacci phase}\label{fibph}
 By combining the phases we have constructed in the preceding subsections, we demonstrate a way to obtain the Fibonacci phase, a topological phase which only has trivial and the Fibonacci anyon as excitations.  \par
Let us first briefly recall the Fibonacci phase. The Fibonacci phase is characterized by $(G_2)_1$ WZNW CFT. The central charge is $c=14/5$ and the primaries are identity $I$ and $\tau$ with conformal weight $2/5$ subject to fusion rule $\tau\times\tau=I+\tau$. \par
 Consider a geometry obtained by putting the networks of the $\nu=6$ IQH islands on the top of the $\nu=2$ IQH networks which yields
the $SU(2)_1\times SU(2)_3$ topological phase with edge mode described by $SU(2)_1\times SU(2)_3$ WZNW CFT. Such a configuration may occur in networks of multi-layer IQH islands.
Since the central charge of the $SU(2)_1\times SU(2)_3$ WZNW CFT is given by $c=1+9/5=14/5$, which is identical to the one of the $(G_2)_1$ WZNW CFT. This
may hint us that the $SU(2)_1\times SU(2)_3$ topological phase is related to the Fibonacci phase. \par 
A key formalism to obtain the Fibonacci phase is anyon condensation~\cite{gils2009collective,Bais2009,ludwig2011two,davydov2013witt,eliens2014diagrammatics} which is a generalization of the vortex proliferation to anyonic system. In the process of the anyon condensation, we proliferate anyons that have the bosonic property in the sense of having integer conformal weight, allowing us to identify the vacuum with the anyons that are condensed.  The reader may also benefit from a relatively nontechnical review \cite{Burnell2018}.  While the ``bosonic" anyons are bosonic in the sense that they braid trivially with each other, they are still anyonic since they braid non-trivially with others. 
In the present context, it can be achieved using the mechanism described in Ref.~[\onlinecite{tsvelik2014integrable}] [see also the discussion around Eq.~(\ref{parafer}) in the previous section]. By this scenario the anyon gas will exist on borders between the islands. 
  Otherwise  not much is known about Hamiltonian formalism to describe the condensation with only few exceptions. For instance, Hamiltonian of the condensation of an anyon in a non-chiral topological phase is given in Ref.~[\onlinecite{bravyi1998quantum}].  The condensation by bosonic anyons put several restrictions on other anyons.   
First, anyons which are related to each other by fusing with the condensed anyons are identified. Second,  a new phase after the condensation admits 
only the anyons that braid trivially with the anyons that are condensed as excitations, 
otherwise they carry a visible non local Dirac string which would cost energy increasing with distance of the separation. More succinctly, in the condensed phase, anyons which braid trivially (non-trivially) with condensing anyons are deconfined (confined).\par
Now we apply this scheme to our case. We carry out condensation of composite anyons in the $SU(2)_1\times SU(2)_3$ topological phase.  
In this phase, there are $2\times 4=8$ types of anyons labeled by $\{I,s\}\times \{I,X,Y,Z\}$. 
In addition to the vacuum, $I\times I$, the $s\times Z$ anyon has integer conformal weight $1$, which can be condensed. Such condensation would be possible by proliferating composite of $h/e$ vortex and $3h/e$ vortices in a $\nu=2$ and $\nu=6$ IQH island.
After condensing this anyon, a new vacuum is $I\times I\simeq s\times Z$, and some anyons are identified by fusing with the $s\times Z$ anyon. For instance, the $s\times X$ and $I\times Y$ anyons are identified as they are related by fusion with the $s\times Z$ anyon. 
Similarly, $I\times Z$ and $s\times I$ are identified as well as $I\times X$ and $s\times Y$. We also need to analyze which anyon remains a deconfined excitation after the condensation. It turns out that the only deconfined excitation is the $s\times X$ anyon. To see this, we have to check that the $s\times X$ anyon has trivial braiding with the $s\times Z$ anyon. We evaluate the  monodromy, a  phase factor obtained by braiding of $s\times X$ and $s\times Z$ anyons. The monodromy is read from the conformal weights of the anyons, which has the form $e^{2\pi i(h_c-h_a-h_b)}$, where $h_a$, $h_b$ and $h_c$ denotes the conformal weight of $s\times X$, $s\times Z$, and $I\times Y=(s\times X)\cdot(s\times Z)$, respectively. From the table.~\ref{labels} and the fact that the conformal weight of the semion $s$ is $1/4$ it follows that the monodromy is trivial ($=1$) implying the $s\times X$ anyon is deconfined. The analogous thought shows that the $I\times Z$ and $I\times X$ anyons are confined. See, table.~\ref{conden}. To summarize, the resulting phase has vacuum and one non-trivial anyon as excitation.  
\begin{table}[htb]
  \begin{tabular}{cc} 
   Anyon & Confined or deconfined\\ \hline
    $s\times X\leftrightarrow I\times Y$& deconfined\\
    $I\times Z\leftrightarrow s\times I$& confined\\
  $I\times X\leftrightarrow s\times Y$& confined
  
  \end{tabular}
    \caption{List of anyons after condensing the $s\times Z$ anyon. The arrow represents identification. Originally the $SU(2)_1\times SU(2)_3$ topological phase has eight anyons which are reduced to two, vacuum and the Fibonacci anyon which is equivalent to the $s\times X$ anyon. }\label{conden}
\end{table}
By noting that the conformal weight of the $s\times X$ anyon is $2/5$ with fusion rule $(s\times X)\cdot(s\times X)=I\times I+I\times Y=I\times I+s\times X$, the resulting phase is the Fibonacci phase~\cite{modular}. Notice that the Fibonacci phase can be obtained by nucleating the $SU(2)_4$ topological liquid~\cite{Mong2014}, or by coupling nucleating anyonic chains~\cite{fib}. \par
In addition to the $SU(2)_1\times SU(2)_3$ topological phase, the $SU(2)_{28}$ topological phase is also associated with the Fibonacci phase as the $SU(2)_{28}$ topological phase
carries central charge $c=14/5$, which coincides with the one of the Fibonacci phase [$(G_2)_1$ WZNW CFT]. Following the similar argument explained above, one can show that the anyonic condensation leads the $SU(2)_{28}$ phase to the Fibonacci phase~\cite{davydov2013witt}.

\subsection{General case}
The generalization of our constructions to other cases of $k$ is straightforward. One can start with networks of IQH islands with filling fraction $\nu=2k$ and introduce the $SU(2)_k$ current-current interaction. Instead of Eq.~(\ref{emb6}), utilizing the following conformal embedding
\begin{equation}
    U(2k)_1=SU(2)_k\oplus U(1)\oplus  SU(k)_2,
\end{equation}
the $SU(2)_k$ topological phase is obtained. Likewise, the conjugate phase becomes the $U(1)\times SU(k)_2$ topological phase. 

\begin{figure}[t]
\subfigure[]{%
		\includegraphics[clip, width=0.25\columnwidth]{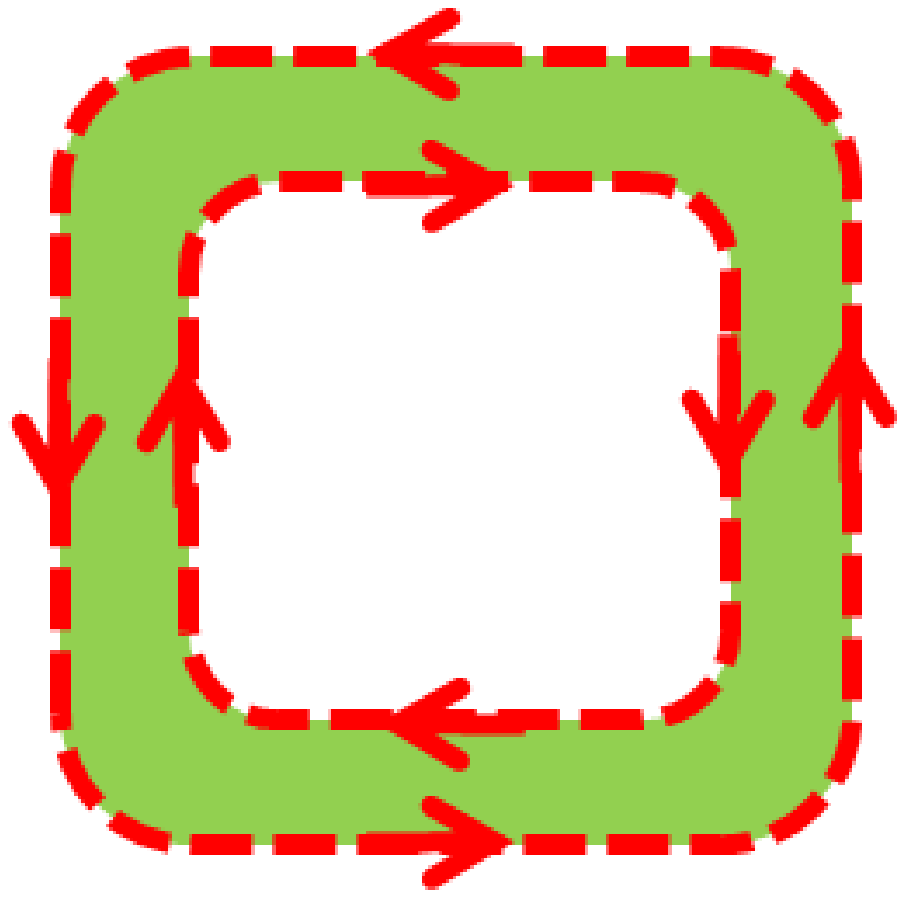}}
\subfigure[]{%
		\includegraphics[clip, width=0.6\columnwidth]{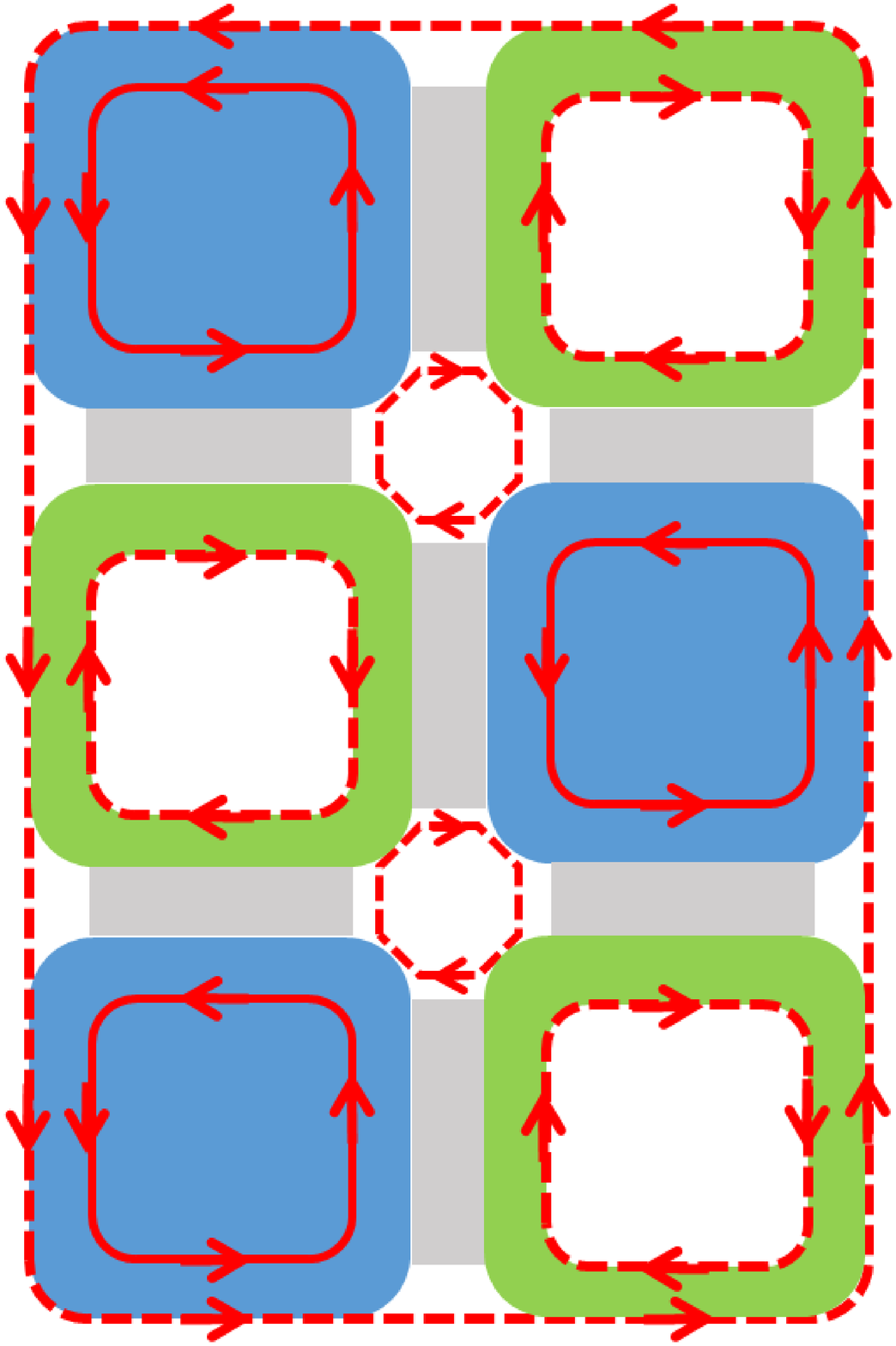}}%
 \caption{(a) A spin island, an insulating island which hosts  a critical chain on its boundary. The criticality is governed by the $SU(2)_k$ WZNW CFT. Counter-propagating gapless modes regarding to this CFT is depicted by dashed red lines.
 (b)An alternative construction of the $SU(2)_k$ topological phase. IQH islands (blue color) and islands each of which has a critical spin chain (green color) are placed in a checkerboard pattern. The red bold line represents edge mode of the $U(1)\times SU(k)_2$ sector whereas the red dashed line is the edge mode of the $SU(2)_k$ sector. Areas marked by grey color between adjacent islands denote interactions described by Eq.~(\ref{cr}).  }
 \label{threecolors}
	
\end{figure}
\section{Alternative construction: Networks of IQH islands and spin chain islands}\label{alt}
In this section, we present an alternative construction of the $SU(2)_k$ topological phase. Now the network consists of IQH and insulating islands arranged in a checkerboard pattern. An insulating island contains a critical spin $S =k/2$ chain on the boundary.

\par
We prepare an IQH island at filling fraction $\nu=2k$ (more precisely, a \textit{spinful}  IQH island with $\nu=2k$ which is realized by a material with a weak Land{\'e} $g$-factor) and the island, that we call spin island, with an inert bulk hosting an integrable  spin-$k/2$ chain on the boundary  [see, Fig.~\ref{threecolors}(a)].  

Defining the $k/2$-spin operator at site $j$ as $\bm{S}_j$, one can write down the most general form of the Hamiltonian of spin $S=k/2$ antiferromagnet with nearest neighbor interactions:
\begin{equation}
    H = J\sum_n P_{k}(\bm{S}_n\bm{S}_{n+1}),
\end{equation}
where $P_k(x)$ is a polynomial of $k$-th degree. The chain becomes integrable for polynomials of special form, namely~\cite{babujian1983exact}
\begin{equation}
    P_k(x) = -\sum_{j=1}^k\Big(\sum_{n=1}^j\frac{1}{n}\Big)\prod_{l=0,l \neq j}^k\frac{x - l(l+1) + k(k+2)/4}{j(j+1) - l(l+1)},
\end{equation}
for instance, up to a prefactor we have 
\begin{equation}
    P_1(x) = x, ~~P_2(x) = x-x^2, ~~ P_3(x) = -x - \frac{8}{27}x^2 +\frac{16}{27}x^3.
\end{equation}
The integrable spin chains are always critical; 
their  long wave length behavior is governed by the $SU(2)_k$ WZNW CFT~\cite{Affleck1987} with the  Hamiltonian 
\begin{equation}
H_{\text{WZNW}}=\frac{2\pi v_s}{k+2}\int dx \sum_{a=1,2,3}(:j^a_{R}j^a_{R}:+:j^a_{L}j^a_{L}:)\label{wzw},
\end{equation}
where $j^a_{R,L}$ are chiral $su(2)_k$ Kac-Moody currents and $v_s$ is the spinon velocity. This is different from the generic situation where only chains with half-integer spins are critical in the $SU(2)_1$ universality class. 


With these two types of islands, we consider the configuration shown in Fig.~\ref{threecolors}(b), where the IQH islands and the spin islands are placed in a checkerboard pattern. The continuum limit of the antiferromagnetic exchange interaction  between adjacent IQH and spin islands is  
\begin{eqnarray}
H=&&\int dx\sum_{\alpha,i}iv\psi^{\dagger}_{R,\alpha,i}\partial_x\psi_{R,\alpha,i}\nonumber\\
&+&\frac{2\pi v_S}{k+2}\sum_{a=1,2,3}(:j^a_{R}j^a_{R}:+:j^a_{L}j^a_{L}:)\nonumber\\
&+&\sum_{a=1,2,3}\lambda_{2k} J_{R}^a j_{L}^a,\label{cr}
\end{eqnarray}
where $J_{R}^a$ is the $SU(2)$ current in the IQH island and $j_{L}^a$ is the spin current of the spin chain in the spin island.
(Note the difference between the upper and lower case latter of the $SU(2)$ current distinguishing the one of the IQH and the spin island.) 
Here we dropped  the interaction of the currents with the same chirality as it is  exactly marginal.
The advantage of this construction is that when we assume that there are only local interactions, the form of interaction [the third term in the r.h.s of Eq. (\ref{cr})] arises naturally and does not require any fine tuning.  
Indeed, since the fermions are chiral there is no backscattering between the IQH edge and the spin chain and the only possible interactions are between the smooth parts of the magnetization described by the spin currents. We have chosen the SU(2)-symmetric form of the interaction, but, in fact, the anisotropy is irrelevant.
The interaction gaps out the $SU(2)_k$ sector on every island (both the spin and the IQH ones) and also confines the chiral gapless modes to their particular islands. The mechanism is the same as the one in the previous section, the only difference lies in the microscopics. 

In addition to kinks in the gapped area between adjacent islands and
$h/e$ vortices in an IQH island, an anyon can bind to a spin excitation in a spin island. In the continuum limit we have  
\begin{eqnarray}
&& S_n^a=j_{R}^a+j_{L}^a+i\text{const}(-1)^n\text{tr}(\sigma^a \phi_{1/2})+\cdots,\label{sping}
\end{eqnarray}
where $\phi_{1/2}$ is the spin-1/2 primary field ($2\times 2$-dimensional matrix field) of the $SU(2)_k$ WZNW model. For $k=1$ we also have 
\begin{equation}
    ({\bm S}_n{\bm S}_{n+1}) = T_R +T_L + \text{const}(-1)^n\text{tr}\phi_{1/2} +\cdots, 
\end{equation}
where $T_{R/L}$ are holomorphic/anti-holomorphic components of the stress-energy tensor and the dots stand for less relevant operators. 
This means that the spin excitation in the spin island hosts an anyon corresponding to the primary field $\phi_{1/2}$ in the $SU(2)_k$ WZNW CFT. For $k=1$ this can be introduced by, for instance, a local variation of the exchange integral giving rise to dimerization resulting in a semion in the KL state.  Excitations related to other primary fields  in the WZNW CFT, such as the $Y$ or $Z$ anyons in the $SU(2)_3$ topological phase are created by other operators. For instance, deviations from the critical point are related to tr$\phi_1$ - the trace of the operator in the adjoint representation:
\begin{equation}
  \delta H = \delta\sum_n  ({\bm S}_n{\bm S}_{n+1}) \sim \delta \int dx  \text{tr}\phi_1 +\cdots. \label{phi2}
\end{equation}
\par
As we have mentioned above, the criticality of the spin chain is destroyed by relevant perturbations. However, we argue that for odd $k$ the topological phase we just described is stable. In what follows, we focus on the $SU(2)_k$ sectors in an IQH and spin islands, omitting the chiral edge modes of the $U(1)\times SU(k)_2$ sectors in an IQH island as they are not affected by interactions that we consider. This means that relevant part of our system (before turning on interactions) is described by the product theory:
\begin{equation}
    \left[ SU(2)_k \right]_R \times \left[ SU(2)_k \right]_L \times \left[ SU(2)_k \right]_R,
    \label{eqn:non-interacting}
\end{equation} 
where the first term corresponds to the $SU(2)_k$ sector in the IQH island and the second and third to the gapless mode of the spin chain.
\par 

\begin{figure}[h]
 \begin{center}
  \includegraphics[width=80mm]{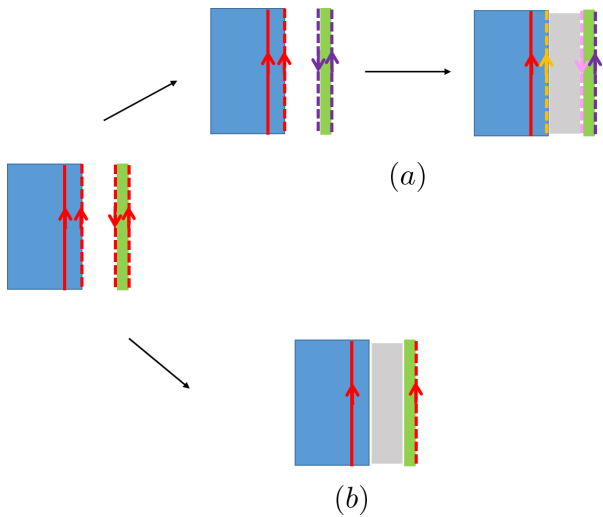}
 \end{center}
 \caption{A schematic picture of edge modes in the vicinity of borders between adjacent IQH and spin islands. Without interaction, the IQH island has edge modes of the $SU(2)_k$ sector (red dashed line) and the $U(1)\times SU(k)_2$ sectors (red bold line), whereas the spin island has counter-propagating edge modes of the $SU(2)_k$ sector, as depicted in the leftmost figure. 
 When  the spin chain in the spin island is perturbed from the $SU(2)_k$ critical point, it flows to the $SU(2)_1$ fixed point (purple dashed lines). As a consequence, the current-current interaction becomes the one involving the right $SU(2)_k$ and left $SU(2)_1$ chiral vector current in $[SU(2)_{k}]_R\otimes [SU(2)_{1}]_L$ WZNW model. This interaction  flows to the non-trivial fixed point described by Eq.~(\ref{achiral}) as shown in (a).
 On the other hand, when we introduce the current-current interaction given in Eq.~(\ref{cr}), a pair of the edge modes of the $SU(2)_k$ sector are gapped out, which is illustrated in (b). The fixed point of~(b) has a lower central charge and hence by the $c$-theorem it cannot flow to the fixed point of (a) with introduction of small perturbations.}\label{RGedge}
 \end{figure}

We know that generically a spin chain with half integer spins is critical in the $SU(2)_1$ universality class. Hence a relevant perturbation of the spin chain such as the one in Eq.~(\ref{phi2}) drives the chain from the $SU(2)_k$ to the $SU(2)_1$ critical point~\cite{haldane1983nonlinear}. 
Then in the infrared  the spin currents $j_{L}$ in Eq.~(\ref{cr}) are transmuted into $SU(2)_1$ currents. Therefore, the current-current interaction between adjacent IQH and spin islands becomes the one involving the right $SU(2)_k$ and left $SU(2)_1$ chiral vector current in $[SU(2)_{k}]_R\otimes [SU(2)_{1}]_L$ WZNW model. 
According to Ref.~[\onlinecite{andrei1998chiral}], the perturbation by such a current-current interaction yields a massless flow towards a non-trivial fixed point preserving chiral central charge which is defined as difference of central charges of the right and left moving sectors, $c_R-c_L$. 
The criticality at this fixed point is shown to be 

\begin{equation}
\left[SU(2)_{k-1}\right]_R\otimes
\left[\frac{SU(2)_{1}\times SU(2)_{k-1}}{SU(2)_{k}}\right]_L \label{achiral},
\end{equation}
leaving the right moving $SU(2)_1$ sector in the spin island untouched. See also Fig.~\ref{RGedge}(a).
The total central charge, which is defined by summation of central charges of the right and left moving sectors of the fixed point of Eq.~(\ref{achiral}) is given by
\begin{equation}
    c_a=\left[\frac{3(k-1)}{k+1}+\left(1-\frac{6}{(k+1)(k+2)}\right)\right]+1.
\end{equation}
 
We compare $c_a$ with $c_b$ which is the total central charge obtained by perturbation of the non-interacting theory [Eq.~(\ref{eqn:non-interacting})] by the $SU(2)_k$ current-current interaction [given in Eq.~(\ref{cr})]
\begin{equation}
    c_b=\frac{3k}{(k+2)},
\end{equation}
corresponding to the fact that the only surviving mode is the right moving $SU(2)_k$ sector in the spin island [Fig.~\ref{RGedge}(b)]. 
One finds that $c_b<c_a$ holds when $k>1$. 
We know that the system flows to the fixed point $b$ with central charge $c_b$ when other relevant interactions are tuned to zero. Thus, very small deviations from this flow would take the system first to a neighbourhood of the fixed point $b$, and then to the fixed point $a$ with central charge $c_a$, assuming that instabilities cause the system to flow to the fixed point $a$. But since $c_b<c_a$, such flows cannot exist for sufficiently small neighbourhoods around the fixed point $b$ (i.e., for sufficiently small initial values of instabilities) by Zamolodchikov's $c$-theorem~\cite{zamolodchikov1986irreversibility,zamolodchikov1987renormalization}. Hence, there is a window of stability around the fiducial flow, and only large enough perturbations can possibly take the system to the fixed point $a$, 
verifying that the topological phase is stable against the perturbations driving the spin chains from the $SU(2)_k$ critical point.\par
Moreover, the theory (b) is chiral and has the minimal central charge among all theories with the same value of the \textit{chiral} central charge $c_R - c_L$ -- a quantity that describes the gravitational anomaly and hence must be preserved along RG flows. Hence the preceding argument also shows that our fixed point is absolutely stable. \par
For even values of $k$, the spin island hosts an integer spin chain, and although interactions may still be tuned to achieve $SU(2)_k$ criticality, the spin chain is generically gapped~\cite{haldane1983nonlinear}. In this case, we require that such a gap be smaller than a gap produced by the current-current interaction in Eq.~(\ref{cr}), in order for the $SU(2)_k$ topological phase to be stable.

\section{Experimental consequences}
\label{sec:experiment}
In this section we comment on the relevance of our proposal for experimental realizations. We first discuss the IQH network.
\begin{figure}[h]
 \begin{center}
  \includegraphics[width=80mm]{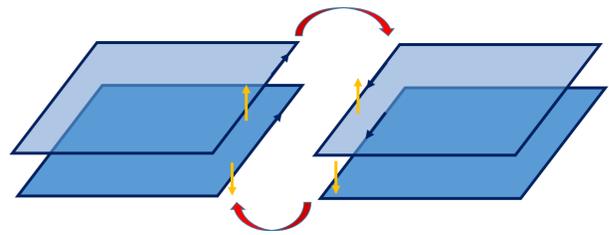}
 \end{center}
 \caption{A pair of $\nu=2$ islands consisting of double layer of $\nu=1$ state. The two edge modes (black lines) in each double layer can be treated as pseudo-spin degrees of freedom~(yellow arrows). 
A tunneling of an electron from one layer between two islands accompanying with backward tunneling of another electron from the second layer (red arrows), which can be regarded as spin backward interaction, is allowed as it preserves momentum and charge in each island. }\label{dl}
 \end{figure}
In the case of $k=1$, the current-current interaction in  Eq.~(\ref{interaction}) can be understood using the bosonization language as a  combination of backward scattering and density-density interaction  (Appendix.~\ref{bosonization}). One possible way to generate  such interaction would be by adjusting a gate voltage, controlling the density of electrons between the islands.

Furthermore, if we introduce networks of double-layer IQH islands,  each island contains two $\nu=1$ states, which can be treated as pseudo-spin degrees of freedom.  Then effective backward pseudo-spin scattering between adjacent islands may occur without breaking  momentum conservation and changing the charge  in each island. In this process, tunneling of an electron from one layer between two islands is compensated by backward tunneling of another electron from the second layer. See also Fig.~\ref{dl}.

The central charge $c$ of the chiral edge modes we discuss here is a topological characteristic of the states. The thermal conductance reflects this topological number.  It is therefore important  to measure it \cite{Cappelli2002,jezouin2013quantum,banerjee2017observed,banerjee2018observation}. The thermal conductance $\kappa$ at temperature $T$ of the edge mode of the topological phase is given by $\kappa=c$ in units of $\kappa_0=\frac{\pi^2k_B^2}{3h}T$. For the  $SU(2)_k$ case we find that the chiral central charge is $c=3k/(k+2)$ and for the conjugate phase it is  $c=2k-3k/(k+2)=k(2k+1)/(k+2)$. 

To  confirm the existence of topological phases further, it would be highly desirable to  probe the anyon statistics. To this date, however, even detection of Abelian statistics remains  a challenging task. Progress has been made by a recent work demonstrating that Aharonov-Bohm oscillations are observed in the $\nu=1/3$ and the $\nu=2/3$ FQH states by experiments of the Fabry-Perot interferometer  \cite{bhattacharyya2019melting,nakamura2019aharonov}, which might provide a good platform for observation of the fractional statistics. In the simplest case of our model, the KL state, one can envisage an interferometer experiment where a $h/e$ vortex is induced in the bulk which gives the semion. Since the edge mode is neutral, thermal conductance or AC conductance would be a useful observable to confirm the statistics. 

For the case where IQH islands are combined with spin islands the exchange interaction is the only possible one, but the obvious problem is how to achieve a tight contact between the Hall islands and the spin chains. We envisage that this can be achieved with the modern methods of molecular beam epitaxy. One may also be inspired by the experiments where topologically protected edge states are brought into contact with microscopic iron clusters~\cite{jack2019observation}.
We can avoid freezing of the spin degrees of freedom in high magnetic fields by lowering the value of the Land\'e $g$-factors, both on the Hall islands and on the spin chains. Nearly isotropic spin $S=1/2$ chains with strongly anisotropic Land\'e $g$-factors are known~\cite{wu2019tomonaga}.\par 

\section{Conclusions} \label{conclusion}
 In this paper, we have constructed the $SU(2)_k$ topological phase and its conjugate phase using networks of interacting IQH islands and spin chains.
 Previous attempts to construct such exotic phases utilized wire construction~\cite{Kane2002,Teo2014,Mong2014,meng2015coupled,Sagi2017,Kane2017,lopes2019non,yang2020topological}, or wire construction emergent from critical anyonic chains~\cite{ludwig2011two,poilblanc2011quantum,fib}.
 The main advantages of our construction is that it gaps out certain degrees of freedom by virtue of the geometry and  
 %
 imposes fewer constraints on the choice of interactions. In other words, to obtain a topological state with a gapped bulk, it is sufficient to generate a gap just in a sub-sector of the system. Then, despite the fact that the interaction does not affect  the chiral modes belonging to the other sectors, these modes remain confined in their islands by virtue of the geometry of the network. 
 
  In our analysis, we have concentrated on the cases of $k=1,2,3$, which are of fundamental importance. The $SU(2)_1$ topological phase is dual to the KL state, one of the spin liquid phases, which has semionic excitations. 
 Previous suggestions for realizations of spin liquid phases include interaction of three nearest-neighbour spins that breaks time reversal symmetry in the form of a mixed product of three spins (see subsequent works, Refs.~\cite{baskaran1989novel,wen1989chiral,peruzzo2014variational} and recent works~\cite{gorohovsky2015chiral,chen2017model,tikhonov2019quantum, ferraz2019spin}).
 Instead of having such interactions, our model contains more attainable  current-current interactions, and breaks time reversal symmetry by the use of IQH islands.
 
We find that the case with $k=2$ is dual to Kitaev's spin liquid phase and the Moore-Read phase both of which have the Ising anyons and half-integer central charge; in our $SU(2)_2$ case $c=3/2$.

The $SU(2)_3$ topological phase that we propose is important as it contains Fibonacci anyons that may be useful for universal topological quantum processing~\cite{Freedman2002}.

 We have proposed a geometry consisting of the $SU(2)_1$ and $SU(2)_3$ topological phases that stabilizes the Fibonacci phase (consisting of the vacuum and Fibonacci anyons only) by employing the scheme of anyon condensation. 
 
 The  phases  we  obtain  are  gapped  in  the  bulk and contain chiral edge modes; we expect therefore that an  introduction  of  perturbations  smaller  than  the  bulk gap will not lead to a phase transition and will not modify the universal properties of the topologically ordered phase. However, in order to obtain these phases we have assumed a specific form of the interaction [spin $SU(2)_k$ current-current interaction]. 
 It is known that a generic form of interaction flows (for $k>1$) either to weak coupling or to strong coupling with the maximal possible symmetry~\cite{balents1996weak,lin1998exact}, in the given case $SU(2k)$,
 which would lead to the $SU(2k)_1$ topological phase with central charge $c=2k-1$ and Abelian anyons in the bulk.  
We therefore consider  an alternative network consisting of IQH islands and spin islands placed in a checkerboard pattern. In this alternative construction, the $SU(2)_k$ current-current interaction arises naturally in view of symmetry and chirality.

\section*{ACKNOWLEDGMENT}
We thank Ofer Aharony, Eyal Leviatan, Adar Sharon, Ryan Thorngren for helpful discussions.
AMT was supported by the U.S. Department of Energy, Office of Basic Energy Sciences, Contract No.~DE-SC0012704. The research at Weizmann was partially supported by the European Unions Horizon 2020 research and innovation programme (grant agreement LEGOTOP No 788715), the DFG
(CRC/Transregio 183, EI 519/7- 1), ISF and MAFAT Quantum Science and Technology grant, an NSF/DMR-BSF 2018643 grant,
the Israel Science Foundation center for Excellence grant (grant number 1989/14), the Minerva foundation with funding from the Federal German Ministry for Education and Research, and the Koshland postdoctoral fellowship.

\appendix
\section{Bosonized form of the action for  \texorpdfstring{$k=1$}{Lg}}\label{bosonization}
As mentioned in the main text, in the case of $k=1$, the decomposition $U(2)=U(1)+SU(2)$ can be understood as the spin-charge separation in the physics of the Tomonaga-Luttinger liquid. The interpretation of this decomposition allows us to rewrite Hamiltonian in Eq.~(\ref{interaction}) in the bosonized form. We will see how gapping out the $SU(2)_1$ sector, which is crucial for a  realization of the KL state, is described  in the  language of bosonization (As a primer for this subject, readers should consult with a standard review such as Ref.~[\onlinecite{senechal2004introduction}]). It turns out that by tuning backward scattering and density-density interaction, one can generate a spin gap preserving the symmetry of the $SU(2)_1$.
Using the bosonization formalism, we will also discuss how the semion arises as a kink in a gapped area between adjacent islands.  \par
To begin with, we introduce four chiral bosonic fields, $\phi_{\alpha}(z), \bar{\phi}_{\alpha}(\bar{z})$, where the subscript of the bosonic fields $\alpha$ takes the value $1$ or $2$ which are interchangeably $\uparrow$ or $\downarrow$, and $\phi_{\alpha} (\bar{\phi}_{\alpha})$ is holomorphic (anti-holomorphic) field which is interpreted as right (left) moving field
with $\partial_{\bar{z}}\phi_{\alpha}=\partial_z\bar{\phi}_{\alpha}=0$. 
Here, we have changed the ($1+1$ dimensional) coordinates from $(t,x)$ to the complex $(z,\bar{z})$ via
$z=-i(x+vt)$, $\bar{z}=i(x-vt)$ with $v$ being velocity of the field.
Using these bosonic fields, the Dirac fields of the edge modes of a $\nu=2$ IQH island are bosonized by
\begin{eqnarray}
\Psi_{R\alpha}&=&\frac{1}{\sqrt{2\pi}}\eta_{\alpha}e^{-i\sqrt{4\pi}\phi_{\alpha}}\nonumber\\
\Psi_{L\alpha}&=&\frac{1}{\sqrt{2\pi}}\bar{\eta}_{\alpha}e^{i\sqrt{4\pi}\bar{\phi}_{\alpha}},
\end{eqnarray}
where $\eta_{\alpha}$ and $\bar{\eta}_{\alpha}$ are Klein factors to ensure the anti-commutation relation between the edge mode with different values of the subscript: $\{\eta_{\alpha},\eta_{\beta}\}=\{\bar{\eta}_{\alpha},\bar{\eta}_{\beta}\}=2\delta_{\alpha,\beta}$, $\{\eta_{\alpha},\bar{\eta_{\beta}}\}=0$.
Non-chiral bosonic fields are defined as
\begin{eqnarray}
\varphi_{\alpha}&=&\phi_{\alpha}+\bar{\phi}_{\alpha}\nonumber\\
\theta_{\alpha}&=&\phi_{\alpha}-\bar{\phi}_{\alpha}
\end{eqnarray}
with commutation relation
\begin{equation}
    [\varphi_{\alpha}(x),\theta_{\beta}(x^{\prime})]=-\frac{i}{2}\text{sgn}(x-x^{\prime})\delta_{\alpha,\beta}.\label{cm}
\end{equation}
These two fields are related via $\partial_x\theta_{\alpha}=-\frac{1}{v}\partial_t\varphi_{\alpha}$.
We also introduce ``charge" and  ``spin" bosonic fields by
\begin{eqnarray}
    \varphi_{c/s}&=&\frac{1}{\sqrt{2}}(\varphi_{\uparrow}\pm\varphi_{\downarrow})\nonumber\\
     \theta_{c/s}&=&\frac{1}{\sqrt{2}}(\theta_{\uparrow}\pm\theta_{\downarrow})\label{spin and charge}
\end{eqnarray}
and similarly for $\phi_{c/s}$, $\bar{\phi}_{c/s}$.
Commutation relation of these charge and spin bosonic fields has the same form as Eq.~(\ref{cm}), $i.e.,$
\begin{equation}
     [\varphi_{A}(x),\theta_{B}(x^{\prime})]=-\frac{i}{2}\text{sgn}(x-x^{\prime})\delta_{A,B}\;\;(A,B=c,s).\label{commutation relation}
\end{equation}
\par
Now we are at the stage of applying the bosonization formalism to the $1+1$-dimensional theory which involves four chiral edge modes between adjacent IQH islands. Using the bosonic fields, the kinetic term in Hamiltonian given in Eq.~(\ref{interaction}) in the main text is rewritten as
\begin{equation}
    \frac{v}{2}\sum_{A=c,s}\int dxdt \Bigl[(\partial_x\theta_A)^2+(\partial_x\varphi_A)^2\Bigr].\label{kinetic}
\end{equation}
Using Eq.~(\ref{current01}),
we introduce following current-current interaction
\begin{equation}
    \lambda_{||}(J_x\bar{J}_x+J_y\bar{J}_y)+\lambda_{\perp}J_z\bar{J}_z,\label{cc}.
\end{equation}
The first two terms in Eq.~(\ref{cc}) is further transformed to
 \begin{equation}
 J_x\bar{J}_x+J_y\bar{J}_y=-\frac{1}{2}\Bigl[(\psi_{\uparrow}^{\dagger}\bar{\psi}_{\uparrow})(\bar{\psi}_{\downarrow}^{\dagger}\psi_{\downarrow})+(\psi_{\downarrow}^{\dagger}\bar{\psi}_{\downarrow})(\bar{\psi}_{\uparrow}^{\dagger}\psi_{\uparrow})\Bigr].\label{xy}
 \end{equation}
The r.h.s of Eq.~(\ref{xy}) is proportional to the back scattering Hamiltonian which is bosonized to
\begin{equation}
\eta_{\uparrow}\bar{\eta}_{\uparrow}\bar{\eta}_{\downarrow}\eta_{\downarrow}\left(\frac{-1}{4\pi^2}\right)\cos\sqrt{8\pi}\varphi_s.\nonumber
\end{equation}
As demonstrated in Ref.~[\onlinecite{senechal2004introduction}], the Hilbert space on which the Klein factors act is characterized by eigenstates of a matrix with eigenvalues $\pm 1$, and one may safely pick up one eigenstate, omitting the rest of the Hilbert space . In the following, we choose the eigenvalue $+1$ and write the back scattering term as 
\begin{equation}
  \left(\frac{-1}{4\pi^2}\right)\cos\sqrt{8\pi}\varphi_s.  \label{xy2}
\end{equation}
The third term in Eq.~(\ref{cc}) becomes
\begin{equation}
J_z\bar{J}_z=\frac{1}{2\pi}(\partial_z\varphi_s)(\partial_{\bar{z}}\varphi_s)=-\frac{1}{8\pi}\Bigl[(\partial_x\theta_s)^2-(\partial_x\varphi_s)^2\Bigr],\label{cz}
\end{equation}
where we have used Eq.~(\ref{curren0}) and the fact that density operator is bosonized as  
\begin{equation}
\psi_{\alpha}^{\dagger}\psi_{\alpha}=\frac{i}{\sqrt{2\pi}}\partial_z\varphi_{\alpha},\;\;\bar{\psi}_{\alpha}^{\dagger}\bar{\psi}_{\alpha}=\frac{-i}{\sqrt{2\pi}}\partial_{\bar{z}}\varphi_{\alpha}.
\end{equation} 
Referring to Eqs.~(\ref{kinetic})(\ref{xy})-(\ref{cz}), it follows that Hamiltonian density regarding to Eq.~(\ref{interaction}) reads 
\begin{eqnarray}
\mathcal{H}_2&=&\frac{v}{2}\Bigl[(\partial_x\theta_s)^2+(\partial_x\varphi_s)^2\Bigr]\label{sg0}\\
&-&\frac{\lambda_{\perp}}{8\pi}\Bigl[(\partial_x\theta_s)^2-(\partial_x\varphi_s)^2\Bigr]-\frac{\lambda_{||}}{4\pi^2}\cos\sqrt{8\pi}\varphi_s. \nonumber
\end{eqnarray}
We have omitted the terms involving $\varphi_c$ as they are decoupled from the ones of $\varphi_s$ and intact by the interaction in Eq.~(\ref{cc}), which is consistent with the fact that we don't admit charge transfer between adjacent IQH islands. By introducing coupling constants as
\begin{equation}
g_1=-\lambda_{||}/(4v),\;
g_2=\lambda_{\perp}/(8v),\label{set}
\end{equation}
and with a little more algebra, one finds that 
Eq.~(\ref{sg0}) is identical to the well-known form of the sine-Gordon Hamiltonian
\begin{equation}
\frac{v_s}{2}\Bigl[\left(\frac{1}{v_s}\partial_t\varphi_s^{\prime}\right)^2+(\partial_x\varphi_s^{\prime})^2\Bigr]+\frac{vg_1}{2\pi^2}\cos\sqrt{8\pi K_s}\varphi^{\prime}_s,\label{scl}
\end{equation}
where
\begin{equation}
v_s=v\sqrt{1-(g_2/\pi)^2},\;\;K_s=\sqrt{\frac{\pi-g_2}{\pi+g_2}},\;\;\varphi^{\prime}_s=\frac{1}{\sqrt{K_s}}\varphi_s\label{sg1}.
\end{equation} 

The renormalization group equation of Eq.~(\ref{scl}) is known to be~\cite{kosterlitz1973ordering}
\begin{equation}
\frac{dK_s}{dl}=-\frac{1}{2\pi^2}K_s^2g_1,\;\;\frac{dg_1}{dl}=-2g_1(K_s-1)\label{rg}
\end{equation}
with $l$ being logarithm scaling factor.
From Eq.~(\ref{rg}), one can find that when $|g_1|>2\pi(K_s-1)$, $g_1$ flows to the strong coupling ($g_1$ is marginally relevant). 
Such a condition can be met when $|g_1|$ is infinitesimal and $K_s<1$, that is, $g_2>0$. 
When $\lambda_{||}=\lambda_{\perp}\equiv \lambda_2$, the spin current-current interaction, Eq.~(\ref{cc}) coincides with the one in Eq.~(\ref{interaction}), and from Eqs.~(\ref{set})-(\ref{sg1}), the condition of the current-current interaction being relevant is  
$\lambda_2>0$.\par
To see that the semion is regarded as a kink in the gapped theory between adjacent IQH islands, we start with Hamiltonian in Eq.~(\ref{scl}) and assume $K_s\simeq 1$ and $g_1>0$:  
\begin{equation}
    \frac{v}{2}\left[(\partial_x\theta_s)^2+(\partial_x\varphi_s)^2\right]+\frac{vg_1}{2\pi^2}\cos\sqrt{8\pi}\varphi_s\label{spinsg}
\end{equation}
In this theory, a compactification radius of the bosonic field, which will play an important role later, is given by $R=1/\sqrt{2\pi}$.
Assuming $g_1$ is large so that $\varphi_s$ is pinned to the minima of the well of cosine potential in Eq.~(\ref{spinsg}), it follows that
\begin{eqnarray}
    \sqrt{8\pi}\varphi_s&=&2\pi N+\pi\;\;(N\in \mathbb{Z})\nonumber\\
    \varphi_s&=&\sqrt{\frac{\pi}{2}}\left(N+\frac{1}{2}\right).
\end{eqnarray}
Since $\varphi_s$ and $\varphi_s+2\pi R$ are identified as identical states, one finds that 
there are two distinct minima, that is, $N=0,1\;(\text{mod}\;2)$, implying that the ground state is two-fold degenerate. This can be intuitively understood as two ground state configurations of spins where all of the spins are up or down. 
With this intuitive interpretation in mind, one naively expects that a spin flip occurring at the interface of domains of the two different spin configurations in the ground state is associated with a kink of the cosine potential in Eq.~(\ref{spinsg}). This anticipation turns out to be correct. To see why, we introduce following spin flip operator which is bosonized to~\cite{witten1994non}
\begin{equation}
    \psi_{R\uparrow}\psi_{L\downarrow}^{\dagger}=\frac{1}{2\pi}e^{-i\sqrt{2\pi}\theta_s}e^{-i\sqrt{2\pi}\varphi_c},\label{flip}
\end{equation}
where $\varphi_c$ is the charge bosonic field defined in Eq.~(\ref{spin and charge}). The vertex operator involving $\varphi_c$ is omitted in the present context as we focus on the spin sector.
The operator in Eq.~(\ref{flip}) behaves as a kink in the spin sector, sending one minima to another of the cosine potential. Indeed, from Eq.~(\ref{commutation relation}), we obtain
\begin{equation}
    e^{i\sqrt{2\pi}\theta_s}\varphi_se^{-i\sqrt{2\pi}\theta_s}=\varphi_s+\sqrt{\frac{\pi}{2}}.\label{kink}
\end{equation}
Furthermore, the operator in Eq.~(\ref{flip}) has conformal weight $(1/4,1/4)$ in the spin sector,
which comes from the fact that a vertex operator $e^{il\theta_s}$
in theory Eq.~(\ref{spinsg}) has conformal weight $(\frac{l^2}{8\pi},\frac{l^2}{8\pi}).$
This concludes that the kink which is bound at the interface of the magnetic domains is identified with the semion.\par

\section{Data of WZNW CFT}\label{data}
Here we list the necessary data for the $SU(2)_k$ and $SU(3)_2$ WZNW CFTs to discuss the anyons in the main text. There are $k+1$ primaries of the $SU(2)_k$ WZNW CFT labeled by $\phi_i$ ($i=0,1/2,\cdots,k/2$) whose conformal weights are given by $i(i+1)/(k+2)$. A relatively elementary introduction to $SU_k(2)$ WZNW models is given in ~\cite{TsvelikBook, james2018non}. The fusion rules of the primary operators  resemble the ones for spins; they are
given by~\cite{DiFrancesco1997}
\begin{equation}
\phi_i\cdot\phi_j=\sum_{\substack{l=|i-j| \\ l-|i-j|\in\mathbb{Z}}}^{\text{min}(i+j,k-(i+j))}\phi_l\label{fusion}.
\end{equation}
In the simplest case, $k=1$ we have only two primary fields, denoted by $\phi_0$ and $\phi_{1/2}$ with conformal weight $0$ and $1/4$, respectively. The non-trivial fusion rule reads as $\phi_{1/2}\cdot \phi_{1/2}=\phi_0$. The $s$ anyon, which is nothing but the semion, corresponds to the primary field $\phi_{1/2}$.\par
When $k=2$, there are three primaries, $I$, $\psi$, and $\sigma$ with conformal weight $0$, $1/2$, and $3/16$. Fusion rules are identical to the ones in the Ising topological phase:
$\psi\times\psi=I$, $\psi\times\sigma=\sigma$, $\sigma\times\sigma=I+\psi$.\par
In the case of $k=3$, there are four primaries, $\phi_0$, $\phi_{1/2}$, $\phi_1$, and $\phi_{3/2}$ corresponding to the $I, X, Y$, and $Z$ anyons in Sec.~\ref{Fib}. The fusion rules are read from Eq.~(\ref{fusion}), giving the Table ~\ref{vortex2}.
\begin{table}[htb]
  \begin{tabular}{c|cccc} 
   & $X$&Y&Z\\ \hline
    $X$&$I+Y$&& \\
    $Y$ & $X+Z$&$I+Y$&\\
    $Z$ &$Y$&$X$&$I$\\
    \end{tabular}
    \caption{Fusion rules of the primary fields in the $SU(2)_3$ WZNW CFT. Fusion rules of a primary field with $I$ are not shown since they are trivial. The missing half of the table may be filled in by commutativity.}\label{vortex2}
\end{table}
\par
With regards to the $SU(3)_2$ WZNW CFT, there are
six primaries labelled by $b_0$, $b_3$, $b_{\bar{3}}$, $b_6$, $b_{\bar{6}}$, $b_8$ with conformal weight $0$, $4/15$, $4/15$, $2/3$, $2/3$, $3/5$, respectively. Their fusion rules are provided in Table.~\ref{su3}. 
\begin{table}[htb]
  \begin{tabular}{c||c|c|c|c|c} 
   & $b_3$& $b_{\bar{3}}$& $b_6$& $b_{\bar{6}}$& $b_8$\\ \hline\hline
    $b_3$&$b_{\bar{3}}+b_6$&&&& \\
    $b_{\bar{3}}$&$b_0+b_8$&$b_3+b_{\bar{6}}$&&&\\
    $b_6$&$b_{\bar{3}}$&$b_3$&$b_{\bar{6}}$&&\\
    $b_{\bar{6}}$&$b_8$&$b_8$&$b_0$&$b_6$&\\
    $b_8$&$b_3+b_{\bar{6}}$&$b_{\bar{3}}+b_6$&$b_{\bar{3}}$&$b_3$&$b_0+b_8$
    \end{tabular}
    \caption{Fusion rules of the primary fields in the $SU(3)_2$ WZNW CFT. }\label{su3}
\end{table}
Particularly, the $b_8$ anyon is of our interest, as it is the Fibonacci anyon.
\begin{table}[t]
  \begin{tabular}{c|c||c|c} 
   Label & Diagram&Label&Diagram\\ \hline
    $I (0)$ & &$b_0 (0)$ &\\
    $X (\frac{3}{20})$ & $\tiny\yng(1)$&$b_3 (\frac{4}{15})$ &$\tiny\yng(1)$\\
    $Y (\frac{2}{5})$ &$\tiny\yng(2)$& $b_{\bar{3}} (\frac{4}{15})$ &$\tiny\yng(1,1)$\\
     $Z (\frac{3}{4})$ &$\tiny\yng(3)$&$b_6 (\frac{2}{3})$ &$\tiny\yng(2)$\\
     &&$b_{\bar{6}} (\frac{2}{3})$&$\tiny\yng(2,2)$\\
    &&  $b_8 (\frac{3}{5})$&$\tiny\yng(2,1)$
  \end{tabular}
    \caption{Young diagrams of four anyons in the $SU(2)_3$ topological phase (left) and those of six anyons in the conjugate phase (right). 
    A number in a parenthesis next to each label indicates conformal weight.  
    }\label{primary2}
\end{table}
\begin{table*}[t]
  \begin{tabular}{c|l||c|l} 
   $r$& Decomposition&$r$&Decomposition\\ \hline
   &&&\\
    $1$ &$\frac{1}{2}=\underbrace{h_{SU(2)_3}(\tiny\yng(1))}_{=3/20}+\Bigl(\frac{1}{12}+\underbrace{h_{SU(3)_2}(\tiny\yng(1))}_{=4/15}\Bigl)$ &$4$ &$2=\underbrace{h_{SU(2)_3}(\tiny\young(\bullet\bullet,\bullet\bullet))}_{=0}+\Bigl(\frac{4}{3}+\underbrace{h_{SU(3)_2}(\tiny\yng(2,2))}_{=2/3}\Bigr)$ \\
   &&&$2=h_{SU(2)_3}(\tiny\young(\bullet\,\,,\bullet))+\Bigl(\frac{4}{3}+h_{SU(3)_2}(\tiny\young(\bullet\,,\bullet,\bullet))\Bigr)$\\ 
   &&&$\;\;\;\;=\underbrace{h_{SU(2)_3}(\tiny\young(\,\,))}_{=2/5}+\Bigl(\frac{4}{3}+\underbrace{h_{SU(3)_2}(\tiny\yng(1))}_{=4/15}\Bigr)$\\\hline
   &&&\\
    $2$ & $1=\underbrace{h_{SU(2)_3}(\tiny\young(\,\,))}_{=2/5}+\Bigl(\frac{1}{3}+\underbrace{h_{SU(3)_2}(\tiny\yng(1,1))}_{=4/15}\Bigr)$&$5$ &$\frac{5}{2}=h_{SU(2)_3}(\tiny\young(\bullet\bullet\,,\bullet\bullet))+\Bigl(\frac{25}{12}+h_{SU(3)_2}(\tiny\young(\bullet\,,\bullet\,,\bullet))\Bigr)$\\
    &$1=\underbrace{h_{SU(2)_3}(\tiny\young(\bullet,\bullet))}_{=0}+\Bigl(\frac{1}{3}+\underbrace{h_{SU(3)_2}(\tiny\yng(2))}_{=2/3}\Bigr)$&&$\;\;\;\;=\underbrace{h_{SU(2)_3}(\tiny\yng(1))}_{=3/20}+\Bigl(\frac{25}{12}+\underbrace{h_{SU(3)_2}(\tiny\yng(1,1))}_{=4/15}\Bigr)$\\
    \hline  
 &&&\\   
    $3$&$\frac{3}{2}=\underbrace{h_{SU(2)_3}(\tiny\yng(3))}_{=3/4}+\Bigl(\frac{3}{4}+\underbrace{h_{SU(3)_2}(\tiny\young(\bullet,\bullet,\bullet))}_{=0}\Bigr)$&$6$ &$3=\underbrace{h_{SU(2)_3}(\tiny\young(\bullet\bullet\bullet,\bullet\bullet\bullet))}_{=0}+\Bigl(3+\underbrace{h_{SU(3)_2}(\tiny\young(\bullet\bullet,\bullet\bullet,\bullet\bullet))}_{=0}\Bigr)$\\
    
&$\frac{3}{2}=h_{SU(2)_3}(\tiny\young(\bullet\,,\bullet))+\Bigl(\frac{3}{4}+h_{SU(3)_2}(\tiny\yng(2,1))\Bigr)$&&\\
&$\;\;\;\;=\underbrace{h_{SU(2)_3}(\tiny\young(\,))}_{=3/20}+\Bigl(\frac{3}{4}+\underbrace{h_{SU(3)_2}(\tiny\yng(2,1))}_{=3/5}\Bigr)$&&    
  \end{tabular}
    \caption{Decomposition of $r$ vortices characterized by conformal weight $r/2$ into the $SU(2)_3$ and the $U(1)\times SU(3)_2$ sectors. The Young tableau with solid dots is excluded due to the fact that two [three]columns of boxes are omitted in the Young diagram representation of a primary in the $SU(2)_3$ [$SU(3)_2$] WZNW CFT.  }\label{decomp}
\end{table*}

\section{Excitations of the \texorpdfstring{$SU(2)_3$}{Lg} topological phase and its conjugate phase}\label{more} In order to elucidate the relation between vortices in a $\nu=6$ IQH island and anyons in the $SU(2)_3$ topological phase, here  
we give a detailed discussion on how an excitation given by $r$ $h/e$ vortices ($1\leq r\leq 6$) in a $\nu=6$ IQH island is decomposed into the ones in the $SU(2)_3$ and the $U(1)\times SU(3)_2$ topological phases.
To this end, it is useful to introduce the Young diagram representation of a primary field of the $SU(2)_3$ and $SU(3)_2$ WZNW CFT, which is summarized in Table.~\ref{primary2}. As a primer to the Young diagram, see, for instance, Ref. [\onlinecite{DiFrancesco1997}].  
A key observation is that $r$ $h/e$ vortices ($1\leq r \leq 6$) in the IQH island have conformal weight $r/2$ which is further decomposed into~\cite{NACULICH1990} 
\begin{equation}
\frac{r}{2}= h_{SU(2)_3}(\Lambda)+\Bigl\{\frac{r^2}{2\cdot 6}+h_{SU(3)_2}
({\Lambda^T})\Bigr\}.\label{su23}
\end{equation}
Here, $h_{SU(2)_3}(\Lambda)$ is a conformal weight of a primary in the $SU(2)_3$ WZNW CFT with a Young tableau $\Lambda$ with $r$ boxes, whereas 
$h_{SU(3)_2}({\Lambda}^T)$ labels a conformal weight of a primary in the $SU(3)_2$ CFT represented by a Young tableau ${\Lambda}^T$ which is obtained by transposing the Young tableau $\Lambda$. The second term in the r.h.s of Eq.~(\ref{su23}) corresponds to a vertex operator of the $U(1)$ sector with charge $re$.
\par
Eq.~(\ref{su23}) can be interpreted in the following way;
the first term in the r.h.s of Eq.~(\ref{su23}) corresponds to the anyon in the $SU(2)_3$ topological phase, whereas the second and third terms in the r.h.s of Eq.~(\ref{su23}) correspond to the anyon in the conjugate phase, namely, the $U(1)\times SU(3)_2$ phase.\par 

The decomposition is not unique in some cases of $r$, which is due to the property of the Young diagram representation of a primary field in the $SU(2)_3$ or $SU(3)_2$ WZNW CFT~\cite{DiFrancesco1997}; two vertical boxes, $\tiny\yng(1,1)$ is omitted in a Young diagram representation of a primary field in the $SU(2)_3$ WZNW CFT, as three vertical boxes $\tiny\yng(1,1,1)$ is in the $SU(3)_2$ WZNW CFT. More succinctly, the fact that ``$\Lambda$ and $\Lambda^T$ are related with each other by transposition" holds up to $\tiny\yng(1,1)$ or $\tiny\yng(1,1,1)$. Furthermore, there is a constraint on the number of columns in Young diagrams corresponding to WZNW primaries. In a Young diagram representation of a primary field in the $SU(2)_3$ $\left[ SU(3)_2 \right]$ CFT, a  column of more than three [two] boxes are not allowed.  
 \par    
To see how the property and the constraint mentioned above works in the decomposition of the vortices, 
consider, as an example, the case of $r=3$. 
There are two possibilities for $\Lambda$, that is, $\Lambda=\tiny\yng(3)$ and $\Lambda=\tiny\yng(2,1)$.
For the first case, we have $\Lambda^T=\tiny\yng(1,1,1)$, accordingly, the decomposition in Eq.~(\ref{su23}) is described by
\begin{equation}
\frac{3}{2}=h_{SU(2)_3}(\yng(3))+\Bigl(\frac{3}{4}+h_{SU(3)_2}(\yng(1,1,1))\Bigr).\label{3/2}
\end{equation}
The Young diagram of the third term on the r.h.s is omitted, thus $\Lambda^T$ is empty diagram, yielding $h_{SU(3)_2}(\Lambda^T)=0$. 
Remembering $h_{SU(2)_3}(\tiny\yng(3))=3/4$ (see table.~\ref{primary2}), the r.h.s of Eq.~(\ref{3/2}) reads as $3/4+3/4+0=3/2$ which verifies the validness of Eq.~(\ref{3/2}). \par 
In the second case, $i.e.,$ $\Lambda=\Lambda^T=\tiny\yng(2,1)$, the decomposition is given by
\begin{equation}
\frac{3}{2}=h_{SU(2)_3}(\yng(2,1))+\Bigl(\frac{3}{4}+h_{SU(3)_2}(\yng(2,1))\Bigr)\label{2,1}.
\end{equation} 
Eq.~(\ref{2,1}) can be rewritten as
\begin{equation}
\frac{3}{2}=h_{SU(2)_3}(\yng(1))+\Bigl(\frac{3}{4}+h_{SU(3)_2}(\yng(2,1))\Bigr),\label{3/22}
\end{equation}
where the two vertical boxes in the diagram in the first term of the r.h.s of Eq.~(\ref{2,1}) is omitted.
Referring to the table.~\ref{primary2}, the decomposition demonstrated in Eqs.~(\ref{3/2}) and (\ref{3/22}) implies that three $h/e$ vortices may have the $Z$ anyon or $X$ anyon in the $SU(2)_3$ topological phase. This is because the second and third terms in the r.h.s of Eqs.~(\ref{3/2}) and (\ref{3/22}) belong to the $U(1)\times SU(3)_2$ sectors, which are suppressed in the $SU(2)_3$ topological phase, allowing us to associate the term on the left hand side to the first term on the r.h.s in Eqs.~(\ref{3/2}) and (\ref{3/22}).\par
We can similarly discuss the decomposition for other cases of $r$, which is summarized in table.~\ref{decomp}. From this table, one finds that the primary field in the $SU(2)_3$ WZNW CFT with the Young diagram representation of $\tiny\yng(2)$ appears in the case of $r=2,4$, implying two or four $h/e$ vortices may bind the $Y$ anyon,  that is, the Fibonacci anyon in the $SU(2)_3$ topological phase. 
\par
We can also argue how an edge excitation of the conjugate phase [\textit{i.e.,} the $U(1)\times SU(3)_2$ topological phase] is understood by multiplication of an anyon in the bulk of the $SU(2)_3$ topological phase and vortices in the large IQH surrounding the networks. For instance, from the case of $r=1$ in Table.~\ref{decomp}, combination of the $X$ anyon in the bulk and a $h/e$ vortex outside the networks corresponds to $b_3$ anyon accompanying charged excitation with conformal weight $1/12$ carrying charge $e$. This argument can be generalized to other cases of $r$. The multiplication of an anyon in the bulk of the $SU(2)_3$ topological phase represented by $r$ boxes of the Young diagram $\Lambda$ and $rh/e$ vortices in the large IQH outside the networks corresponds to the edge excitation which has the form $f_rb_{\Lambda^T}$ in the $U(1)\times SU(3)_2$ sector. Here, $f_r$ is charged excitation characterized by conformal weight $r^2/12$ carrying $re$ charge and $b_{\Lambda^T}$ is  anyonic excitation in the $SU(3)_2$ sector represented by the Young diagram $\Lambda^T$.

\bibliography{ref}

\begin{thebibliography}{79}
\bibitem[{cor()}]{corbino}
\bibinfo{note}{We hope that the following simple example will help the reader
  to understand this fact more clearly. Consider a $p$-wave superconductor in a
  Corbino disk geometry and thread a vortex containing $h/(2e)$ at the center
  of the disk. Both the inner and outer edge modes are influenced by the vortex
  so that Majorana zero modes are propagating along the edges. Let us now
  shrink the inner edge to a point; then we may regard the point as a Majorana
  zero mode bound to the vortex. This picture is closely analogous to the
  bulk-edge correspondence -- one Majorana zero mode is trapped at the center
  of the geometry and another Majorana zero mode is propagating along the outer
  edge.}
\bibitem[{con()}]{conjugatephase}
\bibinfo{note}{It is worth emphasizing that between the setups described in
  Fig.~ \ref{check}(b) and 1(c), the only change is on the edge. The bulk phase
  remains $SU(2)_1$, and is surrounded by an IQH sea in Fig.~ \ref{check}(c)
  (this is in line with the idea that a given edge theory describes not the
  interior itself but the transition between the two gapped phases on either
  side). To actually see the emergence of the conjugate phase in the bulk, it
  might be useful to implement a `folding' trick. The folding trick allows us
  to identify a theory consisting of two topological quantum field theories
  (TQFTs) $A$ and $B$ placed in parallel with the one where the two TQFTs are
  put on top of each other, $A\overline{B}$ (the bar means direction of the
  magnetic field is reversed). Applying this trick to the present case, the
  conjugate phase comprising of $\nu=2$ IQHs networks (the KL state) surrounded
  by a large $\nu=2$ IQH states is mapped to the $SU(2)_1\times
  \overline{U(2)}_1$ topological phase (surrounded by vacuum) which is regarded
  as the $\overline{U(1)}$ topological phase, corresponding to the remaining
  sector of the conformal embedding in Eq.~(1)}.
\bibitem[{mod()}]{modular}
\bibinfo{note}{With the identification presented in Table.~II, and introducing
  characters of the $SU(2)_1\times SU(2)_3$ WZNW CFT as $\chi_i\chi^{\prime}_j$
  ($i=0,1/2$, $j=0,\cdots,3/2$), one can construct the partition function of
  the Fibonacci theory by
  $Z_{(G_2)_1}=|\chi_0\chi^{\prime}_0+\chi_{1/2}\chi^{\prime}_{3/2}|^2+|\chi_0\chi^{\prime}_{1}+\chi_{1/2}\chi^{\prime}_{1/2}|^2$
  which is modular invariant. We also check that the boson $s\times X$ has
  trivial Frobenius-Schur indicator implying the boson is indeed condensable as
  explained in Ref.~\cite{eliens2014diagrammatics}. From the from of the
  modular invariant partition function, $Z_{(G_2)_1}$, the splitting of anyon
  does not occur.}
\end{thebibliography}
\end{document}